\documentclass[aps, prb, twocolumn, superscriptaddress, nofootinbib, longbibliography]{revtex4-2}

\usepackage{graphicx}
\usepackage{dcolumn}
\usepackage{amssymb}
\usepackage{amsmath}
\usepackage{xcolor}
\usepackage{bm}
\usepackage{hyperref}
\hypersetup{
    colorlinks=true,
    linkcolor=blue,
    filecolor=magenta,      
    urlcolor=red,
    citecolor=blue,
}
\usepackage{todonotes}
\usepackage{color}
\usepackage{glossaries}
\usepackage{sidecap} 
\usepackage{orcidlink}
\usepackage{mathtools}

\newacronym{DQMC}{DQMC}{determinant quantum {M}onte {C}arlo}
\newacronym{QMC}{QMC}{quantum {M}onte {C}arlo}
\newacronym{DiagMC}{DiagMC}{diagrammatic {M}onte {C}arlo}
\newacronym{2D}{2D}{two dimensions}
\newacronym{AC}{AC}{analytic continuation}
\newacronym{eph}{$e$-ph}{electron-phonon}
\newacronym{DEAC}{DEAC}{differential evolution for analytic continuation}
\newacronym{DMRG}{DMRG}{density matrix renormalization group}
\newacronym{MEM}{MaxEnt}{maximum entropy}
\newacronym{HMC}{HMC}{hybrid {Monte Carlo}}
\newacronym{QP}{QP}{quasiparticle}
\newacronym{SC}{SC}{superconductivity}
\newacronym{DOS}{DOS}{density of states}
\newacronym{BOW}{BOW}{bond-order-wave}
\newacronym{CDW}{CDW}{charge-density-wave}
\newacronym{HS}{HS}{Hubbard-Stratonovich}
\newacronym{DFT}{DFT}{density functional theory}
\newacronym{XAS}{XAS}{X-ray Absorption Spectroscopy}
\newacronym{FS}{FS}{Fermi surface}
\newacronym{ED}{ED}{exact diagonalization}
\newacronym{1D}{1D}{one dimension}
\newacronym{ARPES}{ARPES}{angle-resolved photoemission spectroscopy}
\newacronym{SSH}{SSH}{Su-Schrieffer-Heeger}

\DeclarePairedDelimiterX{\BraKet}[3]{\langle}{\rangle}{%
  #1 \,\delimsize\vert\, #2 \,\delimsize\vert\, #3%
}

\begin{document}

\title{A DQMC study of the spectral and conductive properties of the two-dimensional Holstein model}

\author{J. Neuhaus\orcidlink{0000-0001-6904-8510}}
\affiliation{Department of Physics and Astronomy, The University of Tennessee, Knoxville, Tennessee 37966, USA\looseness=-1}
\affiliation{Institute of Advanced Materials and Manufacturing, The University of Tennessee, Knoxville, Tennessee 37996, USA\looseness=-1}

\author{B. Cohen-Stead\orcidlink{0000-0002-7915-6280}}
\affiliation{Department of Physics and Astronomy, The University of Tennessee, Knoxville, Tennessee 37966, USA\looseness=-1}
\affiliation{Institute of Advanced Materials and Manufacturing, The University of Tennessee, Knoxville, Tennessee 37996, USA\looseness=-1}

\author{N. Mannella\orcidlink{0000-0002-3490-6041}}
\affiliation{Department of Physics and Astronomy, The University of Tennessee, Knoxville, Tennessee 37966, USA\looseness=-1}
\affiliation{Institute of Advanced Materials and Manufacturing, The University of Tennessee, Knoxville, Tennessee 37996, USA\looseness=-1}

\author{S.~Johnston\orcidlink{0000-0002-2343-0113}}
\affiliation{Department of Physics and Astronomy, The University of Tennessee, Knoxville, Tennessee 37966, USA\looseness=-1}
\affiliation{Institute of Advanced Materials and Manufacturing, The University of Tennessee, Knoxville, Tennessee 37996, USA\looseness=-1}

\date{\today}

\begin{abstract}
We study the dynamical properties of the two-dimensional square lattice Holstein Hamiltonian using numerically exact determinant quantum Monte Carlo simulations. In particular, we report systematic calculations of the model's single-particle spectral function and optical conductivity over a range of phonon energies, electron-phonon coupling strengths, carrier concentrations, and temperatures. In doing so, we map the evolution of the system from a dressed metallic phase to a (bi)polaron liquid/insulator to a charge-density-wave insulator as the carrier concentration is tuned from dilute values to half-filling. We corroborate these results using several equal-time correlation measurements and related proxy quantities reconstructed from time-displaced correlation measurements. This paper is also accompanied by an extensive open data set, covering over 16,640 unique parameter values. 
\end{abstract}

\maketitle


\section{Introduction}\label{sec:intro}

Lattice polarons---composite quasiparticles that arise when local lattice distortions bind to charge carriers---can play an essential role in determining the functional properties of materials~\cite{Alexandrov2010}. For example, polarons have been implicated in materials like conducting polymers~\cite{Bredas2985polarons}, 
tungsten bronzes WO$_{3-x}$~\cite{Schirmer1980Conduction, Salje1994Polarons}, halide perovskites~\cite{Li1999selftrapped, Kandada2020exciton}, oxides like  TiO$_2$~\cite{Moser2013Tunable,Yang2013intrinsic}, SrTiO$_3$~\cite{Frederikse1964electronic, Swartz2018polaronic}, 
colossal magnetoresistive manganites~\cite{Calvani1996Infrared, Allodi2001Ultraslow, Mannella2005nodal, Jooss2007polaron, Mannella2007polaron}, high-temperature superconducting cuprates~\cite{Calvani1996polaronic, Shen2004missing, Shen2007Doping, Mishchenko2008charge} and bismuthates~\cite{Khazraie2018oxygen, Menushenkov2024direct, Naamneh2025persistence}, and beyond~\cite{Franchini2021polarons, dai2025polarons}. Nevertheless, understanding the physics of these systems represents a significant challenge to theory, owing, in part, to difficulties in unbiased modeling at carrier densities relevant to many materials. For example, previous \gls*{ARPES} measurements on correlated materials like  manganites~\cite{Mannella2005nodal, Mannella2007polaron} and high-$T_\mathrm{c}$ cuprates~\cite{Shen2007Doping} have reported broad Gaussian lineshapes. These features have been broadly interpreted in terms of Franck-Condon-like phonon replica bands. While such replicas have been derived in strong coupling expansions for \gls*{eph} coupled models~\cite{Alexandrov1992polaronic, Ranninger1993spectral}, they have yet to be derived using unbiased numerical methods in the adiabatic regime or at large carrier densities relevant to these materials. 

There has been substantial progress in modeling polarons and \gls*{eph} interactions from first principles~\cite{Giustino2017electronphonon, dai2025polarons}. These approaches, however, rely on the Born-Oppenheimer approximation, which is valid in the adiabatic limit and does not account for the quantum-mechanical entanglement between the lattice and electrons. \textit{Ab initio} approaches are thus more suitable for describing static polaron properties rather than dynamical correlation functions needed for comparisons to spectroscopic probes. For this reason, much of our understanding of the spectral properties of polarons has been obtained from studies of low-energy effective models like the Holstein~\cite{Holstein1959Studies}, Fr{\"o}hlich~\cite{Frohlich1954electrons}, and variants of the \gls*{SSH} models~\cite{Barisic1970tightbinding, Su1979solitons, Capone1997small, Sengupta2003Peierls, MalkarugeCosta2023comparative}. In particular, the Holstein model is a frequently adopted, minimal lattice model for \gls*{eph} interactions, and is used as a starting point for nonperturbative numerical studies of small (bi)polarons~\cite{Scalettar1989Competition, Noack1991CDW, Niyaz1993CDW, Ahn1999Spectral, Alder1997Variational, Goodwin2006Greens, Sykora2009Coexistence, Nocera2014interplay, Mishchenko2014diagrammatic, Karakuzu2017superconductivity, Weber2018TwoD, Esterlis2018breakdown, Dee2019temperature, Esterlis2019pseudogap, Wang2020zerotemperature, CohenStead2020langevin, Bradley2021Superconductivity, Nosarzewski2021superconductivity, Araujo2022TwoD, Mitric2022spectral, Kovac2025Signature}.\footnote{Bipolarons form when two polarons bound together; they 
form when the polarons can lower their total energy by sharing the same lattice distortion.} The Holstein model describes the lattice as an Einstein solid and couples the local lattice displacement to the local electron occupancies. Despite its simplicity, the Holstein model hosts rich physics, including \gls*{CDW} order, $s$-wave \gls*{SC}, and small (bi)polarons with large effective masses and self-trapping behavior~\cite{Scalettar1989Competition,Bradley2021Superconductivity,Nosarzewski2021superconductivity}. 

Despite widespread interest in the Holstein model, systematic studies of its spectral properties are broadly lacking. While extensive studies have been possible in the single-particle limit~\cite{Boncia1999holstein, Goodwin2006Greens, Ciuchi1997dynamical, Goodvin2011optical} or for higher carrier concentrations in \gls*{1D}~\cite{Hohenadler2005photoemission, Nocera2014interplay, Mitric2022spectral, Kovac2025Signature}, unbiased calculations for the model's spectral properties in higher dimensions have been challenging. 
For example, spectral functions at large filling fractions have been computed within self-consistent Migdal theory in \gls*{2D}~\cite{Nosarzewski2021spectral}. Conversely, nonperturbative simulations have been conducted for the \gls*{1D} Holstein model in the dilute limit using the hierarchical equations of motion approach~\cite{Mitric2022spectral} and \gls*{DMRG} calculations~\cite{Nocera2014interplay, Kovac2025Signature}, as well as for the \gls*{1D} spinless Holstein model at higher carrier densities using \gls*{QMC}~\cite{Hohenadler2005photoemission}. Spectral functions have also been calculated for the \gls*{2D} Hubbard-Holstein model using \gls*{DQMC}~\cite{Nowadnick2017doping} but were restricted to the high-temperature Mott regime, where the Hubbard repulsion is much larger than the strength of the \gls*{eph} coupling, due to the fermion sign problem. To date, systematic nonperturbative calculations of the spectral functions and optical conductivity for the \gls*{2D} Holstein model are relatively sparse. 

Here we address this open problem by systematically exploring the dynamical and transport properties of the \gls*{2D} Holstein model using \gls*{DQMC}. \gls*{DQMC} is a finite-temperature, numerically exact method, and sign-problem-free for the Holstein model. Leveraging \gls*{HMC} updates with exact Fourier acceleration to reduce autocorrelation times, we can overcome the long autocorrelation times that often hinder traditional \gls*{QMC} methods that rely on local updates for sampling~\cite{SmoQyDQMC1, SmoQyDQMC2, Ostmeyer2025Minimal}. Real-axis dynamical correlation functions are reconstructed using the \gls*{DEAC}~\cite{Neuhaus2024SmoQyDEAC1, Neuhaus2024SmoQyDEAC2} and \gls*{MEM}~\cite{Jarrell1996Bayesian} methods for fermionic and bosonic correlation functions, respectively. Furthermore, by jointly analyzing measurements of the single-particle density of states at the Fermi energy, the dc conductivity, the uniform spin structure factor, and the charge structure factor at the \gls*{CDW} ordering wavevector $\boldsymbol{Q}=(\pi,\pi)$, we characterize the properties of the \gls*{2D} Holstein model across a broad range of temperatures, carrier concentrations, phonon energies, and \gls*{eph} coupling strengths. Our analysis uncovers regions of (bi)polaron conductivity, as well as regions where bipolarons freeze into a disordered, phase-separated state with domains of short-range \gls*{CDW} correlations. Our results have implications for understanding small bipolaronic systems at high carrier concentrations, and for determining what factors might limit \gls*{SC} in this model~\cite{Esterlis2018breakdown}.

\section{Model and Methods}\label{sec:methods}
\subsection{The Holstein Hamiltonian}
We study the \gls*{2D} Holstein model~\cite{Holstein1959Studies} defined on an $N = L\times L$ square lattice with lattice constant $a$ and periodic boundary conditions. The model's Hamiltonian is 
\begin{equation}\label{eq:HolsteinHamiltonian}
    \begin{split}
\hat{H}= & -t\sum_{\langle i,j\rangle,\sigma}\hat{c}_{i,\sigma}^{\dagger}\hat{c}_{j,\sigma}^{\phantom{\dagger}}-\mu\sum_{i,\sigma}\hat{n}_{i,\sigma}\\
 & +\sum_{i}\left(\frac{1}{2M}\hat{P}_{i}^{2}+\frac{M\Omega^{2}}{2}\hat{X}_{i}^{2}\right)\\
 & +\alpha\sum_{i,\sigma}\hat{X}_{i}\left(\hat{n}_{i,\sigma}-\frac{1}{2}\right).
\end{split}
\end{equation}
Here $\hat{c}_{i,\sigma}^{\dagger}$ ($ \hat{c}_{i,\sigma}^{\phantom{\dagger}}$) creates (annihilates) a spin-$\sigma$ electron on site $i$ with $\hat{n}_{i,\sigma}=\hat{c}_{i,\sigma}^{\dagger}\hat{c}_{i,\sigma}^{\phantom{\dagger}}$ the corresponding number operator; $t$ is the nearest-neighbor hopping integral and $\mu$ is the chemical potential; $\langle i,j\rangle$ denotes a sum over nearest-neighbors; $\hat{X}_i$ ($\hat{P}_i$) is the displacement (momentum) operator for the oscillator at site $i$ with $M$ and $\Omega$ the associated phonon mass and frequency, respectively; and $\alpha$ parametrizes the strength of the \gls*{eph} coupling.

Throughout, we set $\hbar=k_\text{B}=t=M=a=1$ and report our results in terms of a dimensionless \gls*{eph} coupling strength $\lambda = \alpha^2/(M\Omega^2 W)$, where $W=8t$ is the noninteracting electronic bandwidth in \gls*{2D}. 

\subsection{Determinant Quantum Monte Carlo}
We solved Eq.~\eqref{eq:HolsteinHamiltonian} using numerically exact \gls*{DQMC}~\cite{White1989numerical}, as implemented in the \texttt{SmoQyDQMC.jl} package~\cite{SmoQyDQMC1, SmoQyDQMC2}. Our implementation utilizes a combination of \gls*{HMC} updates as well as reflection and swap updates to reduce autocorrelation times, which allows us to simulate phonon energies in the challenging adiabatic regime $\Omega/W \ll 1$. 

Unless otherwise specified, we carried out all simulations on $L=14$ lattices with four walkers, each with $5\times 10^3$ burn-in sweeps followed by $2\times 10^3$ measurement sweeps, divided into 100 bins, yielding 400 bins in total. During the simulation, we used  \texttt{SmoQyDQMC.jl}'s built-in chemical potential tuner~\cite{Miles2022dynamical} to maintain the targeted electronic density. We carried out systematic simulations over 16,640 different parameter combinations, as detailed in Appendix~\ref{AppendixParameters}. We also generated various single-particle electron spectral functions and optical conductivity curves via \gls*{AC} as detailed in the same appendix and Sec.~\ref{sec:AC_methods}. The entire dataset has been posted to an open data repository (see Data Availability). In the following sections, we present and discuss a representative subset of the results. 

\subsection{Measurements}
To determine singlet formation indicative of bipolaron formation, and assess whether the system is in a \gls*{CDW} phase, we measure the strength of the spin and charge correlations. We measured the spin $S_\mathrm{s}(\boldsymbol{q})$ and charge $S_\mathrm{c}(\boldsymbol{q})$ structure factors, which are obtained from the equal-time real-space spin-spin 
\begin{equation}\label{eq:Sr}
    C_\text{s}(\boldsymbol{r}_i)=\frac{1}{N}\sum_{j} \left\langle \hat{S}_z({\boldsymbol r}_i+{\boldsymbol{r}}_j)\hat{S}_z({\boldsymbol r}_j)\right\rangle
\end{equation}
and density-density 
\begin{equation}
    C_\text{c}(\boldsymbol{r}_i)=\frac{1}{N}\sum_{j} \left\langle \hat{n}({\boldsymbol r}_i+{\boldsymbol{r}_j})\hat{n}({\boldsymbol r}_j)\right\rangle,
\end{equation}
correlation functions, respectively. Here  
$\hat{S}_z(\boldsymbol{r}_i) = \frac{1}{2}(n_{i,\uparrow}-n_{i,\downarrow})$ is the $z$-axis projection of the local spin operator and $\hat{n}({\boldsymbol r}_i) = \sum_\sigma n_{i,\sigma}$ is the local density operator. 
The corresponding momentum-space structure factors are then obtained via the Fourier transform 
\begin{equation}
    S_{\mathrm{s}/\mathrm{c}}(\boldsymbol{q})=\sum_{j}e^{-\mathrm{i}\boldsymbol{q}\cdot\boldsymbol{r}_j}C_{\mathrm{s}/\mathrm{c}}({\boldsymbol{r}_j}).
\end{equation}

The half-filled ($\langle n\rangle = \frac{1}{N}\sum_{i,\sigma} \langle \hat{n}_{i,\sigma}\rangle=1$) square lattice Holstein model has a well-documented finite temperature phase transition to an insulating \gls*{CDW} phase with ordering wave-vector $\boldsymbol{Q} =(\pi, \pi)$ in which electrons randomly localize on one of the two equivalent sub-lattices. The ordered phase may also be viewed as a frozen bipolaron lattice at strong coupling, in which bipolarons preferentially localize on a single sublattice. This transition is signaled by the extensive growth of $S_\text{c}(\boldsymbol{Q})$ at the transition temperature (see, for example, Refs.~\cite{Marsiglio1990pairing, Dee2019temperature, Bradley2021Superconductivity,Weber2018TwoD} and references therein). When doped away from half filling, \gls*{DQMC} simulations in the strong coupling limit find that $S_\text{c}(\boldsymbol{q})$ retains a broad peak centered at $\boldsymbol{Q}$, indicating that short-range remnants of the \gls*{CDW} charge order are present in the system~\cite{Esterlis2018breakdown, Nosarzewski2021superconductivity, Issa2025learning}. Close to half-filling, the system is also prone to phase-separate into small domains of \gls*{CDW} order surrounded by electron depleted regions, particularly in the adiabatic regime~\cite{Esterlis2018breakdown, Bradley2021Superconductivity, Cheng2023Machine}. Thus, tracking the $S_\text{c}(\boldsymbol{Q})$ peak as a function of doping and temperature indirectly probes the formation and subsequent clustering of bipolarons. Additionally, suppression of the uniform component of the spin structure factor $S_{s}(\boldsymbol{q}=0)$ indicates singlet formation in the system and, by extension, provides another marker of bipolaron formation. 

We also measured several imaginary time and dynamical response functions to assess the model's transport and spectroscopic properties. The single-particle spectral function $A(\boldsymbol{k},\omega)$ is obtained by \gls*{AC} (see below) of the momentum space single-particle electron Green's function
\begin{equation}\label{eq:Green}
    G_\sigma(\boldsymbol{k},\tau) = \frac{1}{N}\sum_{i,j}e^{-{\rm i}\boldsymbol{k}\cdot(\boldsymbol{r}_i-\boldsymbol{r}_j)}\left\langle \hat{c}^{\phantom{\dagger}}_{i,\sigma} (\tau) \hat{c}^\dagger_{j,\sigma}(0) \right\rangle,
\end{equation}
for $\tau \in [0, \beta)$. The spectral function is then obtained from inverting the integral equation 
\begin{equation}\label{eq:ACfermion}
    G_\sigma(\boldsymbol{k}, \tau) = \int_{-\infty}^{\infty}d\omega\frac{e^{-\tau\omega}}{1+e^{-\beta\omega}}A_\sigma(\boldsymbol{k}, \omega). 
\end{equation}
Similarly, we obtain the \gls*{DOS} per spin species by performing \gls*{AC} on the local Green's function $G_\sigma(\boldsymbol{r}=0, \tau) = \frac{1}{N}\sum_{\boldsymbol{k}} G_\sigma(\boldsymbol{k},\tau)$. 

To assess the model's transport properties, we measured the current-current correlation function  
\begin{equation}\label{eq:currentcurrent}
    \Lambda_{\alpha,\beta}\left(\boldsymbol{r}_i,\tau\right)=\frac{1}{N}\sum_{j}\left\langle \hat{J}^\alpha(\boldsymbol{r}_i+\boldsymbol{r}_j,\tau)\hat{J}^\beta(\boldsymbol{r}_j,0)\right\rangle,
\end{equation}
where
\begin{equation}\label{eq:current}
    \hat{J}^\alpha(\boldsymbol{r}_i)=-\mathrm{i}t\sum_\sigma \left(c_{i+\alpha,\sigma}^{\dagger}c_{i,\sigma}^{\phantom{\dagger}}-c_{i,\sigma}^{\dagger}c_{i+\alpha,\sigma}^{\phantom{\dagger}}\right)
\end{equation}
is the current operator.
Here, the index $i+\alpha$ denotes the unit cell shifted in the $\alpha \in \{x,y\}$ direction from site $i$. Throughout this work, we report the average of the measurements taken along the two crystallographic directions of the lattice $\Lambda = \frac{1}{2}\left[\Lambda_{x,x}\left(\boldsymbol{r}_i,\tau\right)+\Lambda_{y,y}\left(\boldsymbol{r}_i,\tau\right)\right]$. We then extract the optical conductivity $\sigma(\omega)$ by performing \gls*{AC} on $\Lambda(\boldsymbol{q}=0,\tau)$ using a bosonic kernel 
\begin{equation}\label{eq:ACboson}
    \Lambda(\boldsymbol{q}=0, \tau) = \int_{0}^{\infty}d\omega\frac{\omega\left( e^{-\tau\omega}+e^{-(\beta-\tau)\omega}\right)}{1-e^{-\beta\omega}}\sigma(\omega). 
\end{equation}

Finally, we compute proxy measurements for the \gls*{DOS} at the Fermi level $N(\omega = 0)$ and dc conductivity $\sigma_\text{dc}\equiv \sigma(\omega = 0)$ based on imaginary-time displaced correlation measurements~\cite{Schattner2016Ising, Scalettar1999Quantum}
\begin{equation}
    N(0) \approx \frac{1}{2}\sum_\sigma\frac{\beta}{\pi}G_\sigma({\boldsymbol{r}}=0,\tau=\beta/2) \label{eq:N0} 
\end{equation}
and 
\begin{equation}
    \sigma(\omega = 0) \approx \frac{\beta^2}{\pi}\Lambda({\boldsymbol{q}}=0,\tau=\beta/2)  \label{eq:sigma_dc}.
\end{equation}
These approximations become exact in the limit $T\rightarrow 0$. 

\subsection{Analytic continuation}\label{sec:AC_methods}
Inverting Eqs.~\eqref{eq:ACfermion} and \eqref{eq:ACboson} is an ill-posed problem, particularly for noisy imaginary-time quantities like those obtained in a \gls*{DQMC} simulation. Because of this, several numerical methods have been developed to perform \gls*{AC} on noisy data (see Refs.~
\cite{Jarrell1996Bayesian, Nichols2022Parameter, Shao2023progress} and references therein). In this work, 
we primarily employ the \gls*{DEAC} algorithm~\cite{Nichols2022Parameter} as implemented in the \texttt{SmoQyDEAC.jl} package~\cite{Neuhaus2024SmoQyDEAC1, Neuhaus2024SmoQyDEAC2} for \gls*{AC} of fermionic correlation functions [Eq.~\eqref{eq:ACfermion}] due to its superior ability to resolve sharp spectral features compared to methods like \gls*{MEM}. All \gls*{DEAC} calculations were performed using a fitness target of $\chi^2 = 1.0$ [see Eq.~\eqref{eq:chi2}] with a maximum number of iterations of $1\times 10^4$ per run. The full binned data and covariance matrix were utilized for our calculations. The real frequency grid for all calculations was a uniform grid of 601 values on the interval $\omega \in [-15t,15t]$. 

We have found that the \gls*{DEAC} algorithm performs poorly for bosonic correlation functions at low-energy~\cite{Neuhaus2024SmoQyDEAC1}, which makes it ill-suited for calculating the dc and infrared optical conductivities. We therefore used the \gls*{MEM}~\cite{Jarrell1996Bayesian} algorithm to perform the bosonic \gls*{AC} calculations [Eq.~\eqref{eq:ACboson}]. \gls*{MEM} maximizes the posterior probability by minimizing the function 
\begin{equation}
Q=\chi^2/2-\alpha S, 
\end{equation}
where 
\begin{equation}\label{eq:chi2}
    \chi^2=\sum_{i=1}^{N_\tau}\left[\frac{\bar{\Lambda}(\tau_i)-{\Lambda}(\tau_i) }{\sigma_\Lambda(\tau_i)}\right]^2
\end{equation}
and 
\begin{equation}
    S=\int_{-\infty}^\infty d\omega\left(\sigma(\omega)-M(\omega)-\sigma(\omega)\log\left[\frac{\sigma(\omega)}{M(\omega)} \right] \right).
\end{equation}
Here, $\bar{\Lambda}(\tau_i)$ and $\sigma_\Lambda(\tau_i)$ are the average value of the correlation function and associated error obtained from the \gls*{DQMC} simulations, respectively, which have been rotated into the diagonal basis of the covariance matrix. $M(\omega)$ is a ``default'' model function provided by the user, which provides additional information about the known properties of the spectral function. 
To minimize bias in the results, we used a flat default model normalized so that ~\cite{Huang2019Strange}
\begin{equation}
2\int_0^{\omega_c} M(\omega)d\omega=\int_0^\beta\Lambda(\tau)d\tau,     
\end{equation}
where $\omega_c/t = 20$ is a high-energy cutoff. Our \gls*{MEM} calculations were performed using the \texttt{ana\_cont} package~\cite{Kaufmann2023anacont}. In all cases, the $\alpha$ values were optimized using the $\chi^2$-kink method~\cite{Bergeron2016Algorithms}, which finds the value that best balances reducing $\chi^2$ without overfitting against noise. 
The real frequency grid for all \gls*{MEM} calculations was taken to be a uniform grid of 1001 values spanning $\omega/t \in [0,20]$. 

\section{Results}\label{sec:results}
\subsection{Half-filling, $\langle n \rangle = 1$}
\begin{figure}[t]
    \centering
    \includegraphics[width=\linewidth]{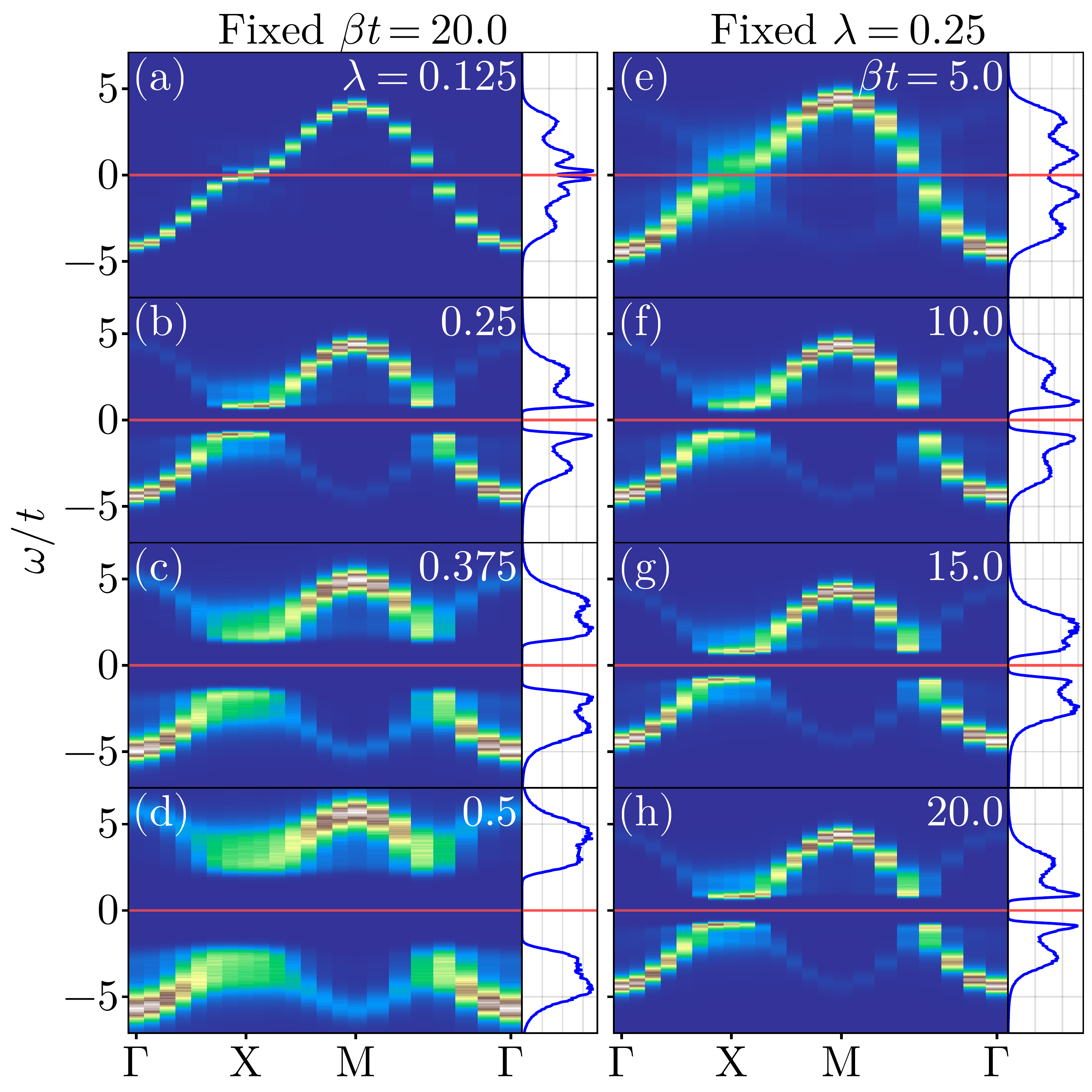}
    \caption{Single-particle spectral functions $A(\boldsymbol{k},\omega)$ for the half-filled ($\langle n\rangle=1$) Holstein model with $\Omega/t=0.5$. Panels (a)-(d) show results for fixed $\beta t=20$ and select $\lambda$ values. Panels (e)-(h) show results for fixed $\lambda=0.25$ and select inverse temperatures $\beta$. The panels to the right of each spectral function show the density of states $N(\omega)$. }
    \label{fig:SpecHalf}
\end{figure}

We begin our analysis by considering the half-filled model with phonon energy $\Omega/t = 0.5$. 
Fig.~\ref{fig:SpecHalf} presents the single-particle spectral function for several values of $\lambda$ and inverse temperatures $\beta$. As already discussed, the Holstein model undergoes a transition to a $\boldsymbol{Q} = (\pi,\pi)$ \gls*{CDW} insulating phase at low temperatures~\cite{Marsiglio1990pairing, Weber2018TwoD, Bradley2021Superconductivity}. In particular, this transition occurs for all values of $\lambda$ at half-filling for sufficiently low temperatures due to the combination of a Van Hove singularity at the Fermi surface and perfect nesting of the Fermi surface.

\begin{figure}[t]
    \centering
    \includegraphics[width=\linewidth]{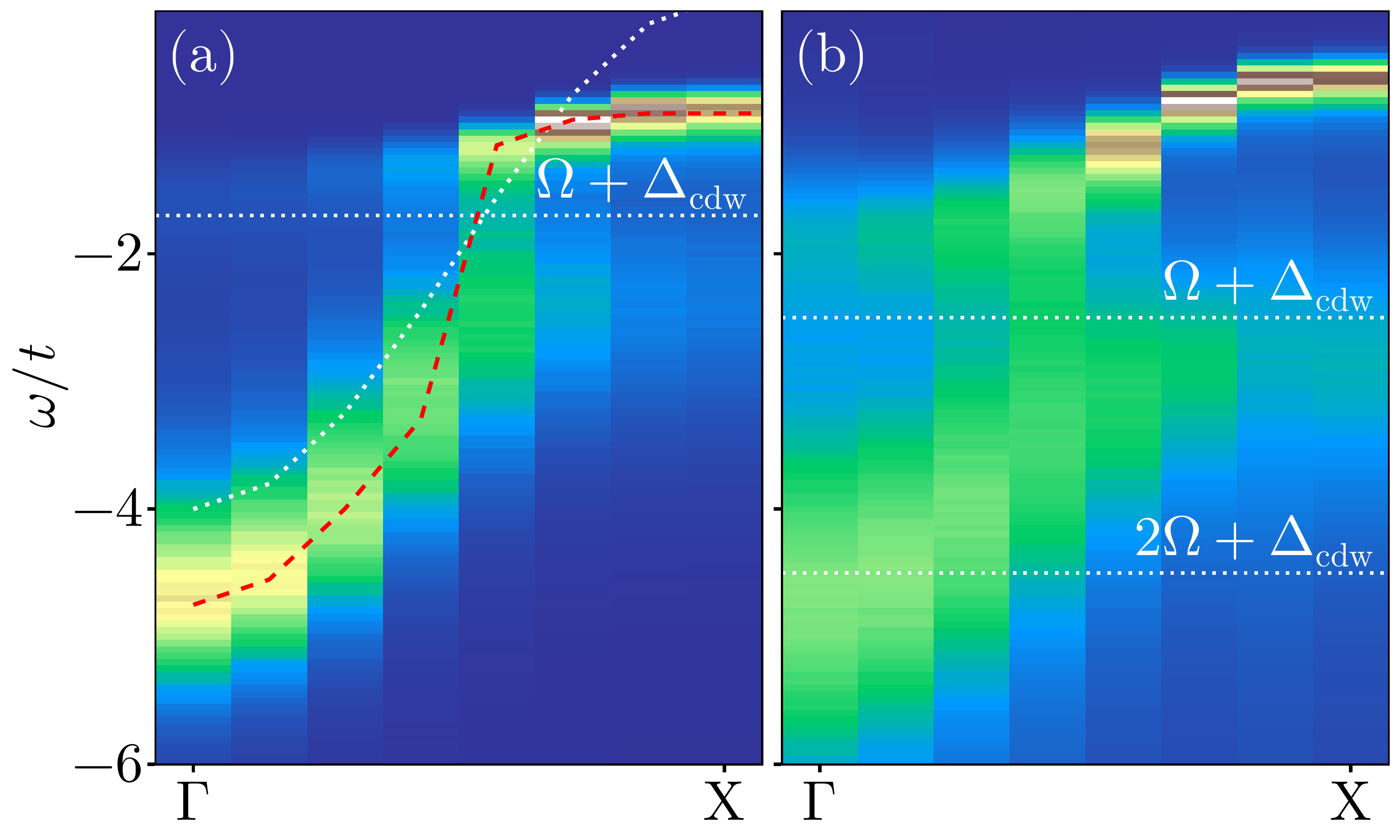}
    \caption{Examples of the Engelsberg-Schrieffer band kinks. (a) Results along the $\Gamma-X$ for $\lambda=0.3,$ $\Omega/t=1$, $\beta t=20$, and $\langle n\rangle=1.0$. The dotted white line plots the noninteracting dispersion. The dashed red line is the observed kinked dispersion. (b) Results along the $\Gamma-X$ for $\lambda=0.3,$ $\Omega/t=2$, $\beta t=20$, and $\langle n\rangle=1.0$ plotted on a square root scale to highlight the spectral weight deeper in the band.
    }
    \label{fig:KinkCondon}
\end{figure}

Our results are consistent with this picture. Figs.~\ref{fig:SpecHalf}(a)-(d) show the evolution of the spectral function at low temperature ($\beta t= 20$) as a function of $\lambda$. For $\lambda = 0.125$, the spectral function largely resembles that of the noninteracting band, but with a very small gap at the Fermi level at the $X$-point. The single-particle density of states $N(\omega)$, shown in the right inset of panel (a), also shows a suppression of spectral weight at the Fermi level, consistent with the presence of a small or partial \gls*{CDW} gap for these parameters. Upon increasing the coupling, we obtain clearer signs of \gls*{CDW} order; already for $\lambda = 0.25$, we can resolve a gap at the Fermi level along with indications of back-folded spectral weight associated with the formation of long-range $\boldsymbol{Q} = (\pi,\pi)$ \gls*{CDW} order. The backfolding is most evident in shadow bands appearing near $\omega/t=5$ at the $\Gamma$-point and $\omega/t = -5$ at the $M$-point. For $\lambda > 0.25$, we also observe the typical Engelsberg-Schrieffer renormalization~\cite{Engelsberg1963coupled} of the band dispersion at higher binding energies, namely ``kinks'' in the band dispersion at $\Delta_\mathrm{cdw} + \Omega$, where $\Delta_\mathrm{cdw}$ is the \gls*{CDW} gap. (A zoom-in on representative kinked dispersions is shown in Fig.~\ref{fig:KinkCondon}.) Note, in all cases shown, we do not observe any clear higher-order features at multiples of the phonon energy.

\begin{figure}[t]
    \centering
    \includegraphics[width=\linewidth]{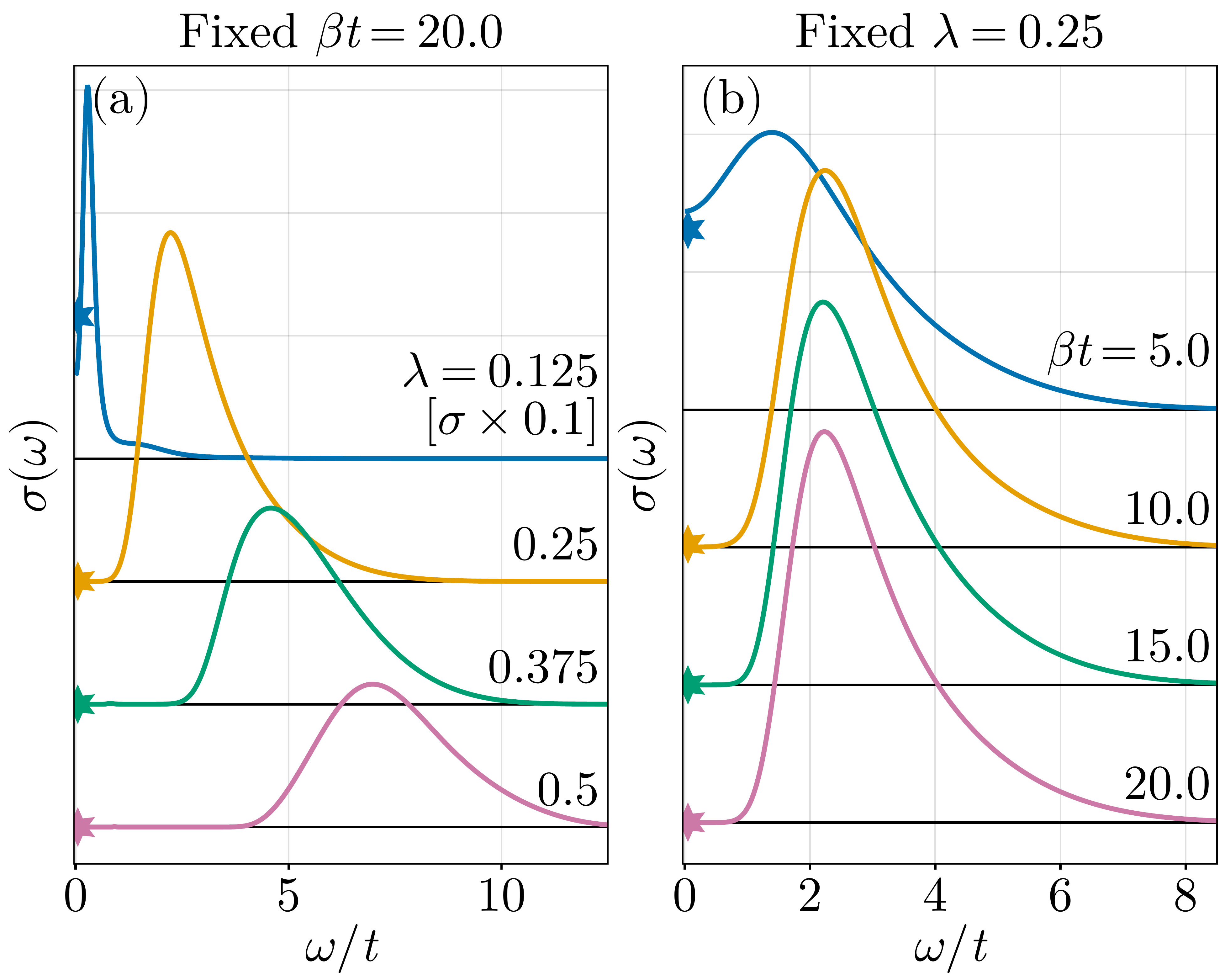}
    \caption{Optical conductivity for the half-filled ($\langle n\rangle=1$) Holstein model with $\Omega/t=0.5$. The star markers placed at $\omega/t=0$ are the dc conductivity values obtained using the imaginary time proxy $\sigma_\mathrm{dc}$ defined in Eq.~\eqref{eq:sigma_dc}. Panel (a) shows results for fixed inverse temperatures $\beta t = 20$ and select $\lambda$, while panel (b) shows results for fixed $\lambda = 0.25$ and select $\beta$. 
    }
    \label{fig:CondHalf}
\end{figure}

Figures~\ref{fig:SpecHalf}(e)-(h) show the evolution of the spectra for a fixed $\lambda=0.25$ as a function of temperature. We already see indications of suppressed spectral weight at the Fermi level and spectral back-folding features at the highest displayed temperature $\beta t = 5$. However, the \gls*{CDW} gap has not yet fully formed at this temperature, consistent with previous reports~\cite{Weber2018TwoD}. The \gls*{CDW} gap opens as the temperature is further reduced and is clearly visible at $\beta t \ge 10$.

Figure~\ref{fig:CondHalf} plots the optical conductivity for the same parameters used in Fig.~\ref{fig:SpecHalf}. In Fig.~\ref{fig:CondHalf}(a), we show results again for fixed $\beta t=20$ as a function of $\lambda$, and have rescaled the weak coupling ($\lambda = 0.125$) result by a factor of $0.1$ for ease of presentation. The stars at $\omega/t = 0$ are the value of the dc conductivity derived from Eq.~\eqref{eq:sigma_dc}, which provides a check on the optical conductivity obtained from \gls*{AC}. The spectrum at weak coupling has a nonzero dc conductivity and a secondary peak at an energy $\omega/t \approx 0.3$, just below the phonon energy $\Omega$. This structure is consistent with predictions derived from Migdal-Eliashberg theory~\cite{Marsiglio1995signatures} and suggests that the system is metallic for these parameters. Increasing the \gls*{eph} coupling to $\lambda=0.25$ completely suppresses the Drude response, consistent with the formation of a fully gapped \gls*{CDW} order. In these cases, the spectral weight is transferred into a peak at finite energy, which corresponds to transitions across the \gls*{CDW} gap. For $\lambda = 0.25$, this peak is asymmetric but becomes more symmetric with increasing \gls*{eph} coupling, which we attribute to the self-energy broadening of the electronic structure [see Fig. \ref{fig:SpecHalf}(d)]. 

\begin{figure}[t]
    \centering
    \includegraphics[width=\linewidth]{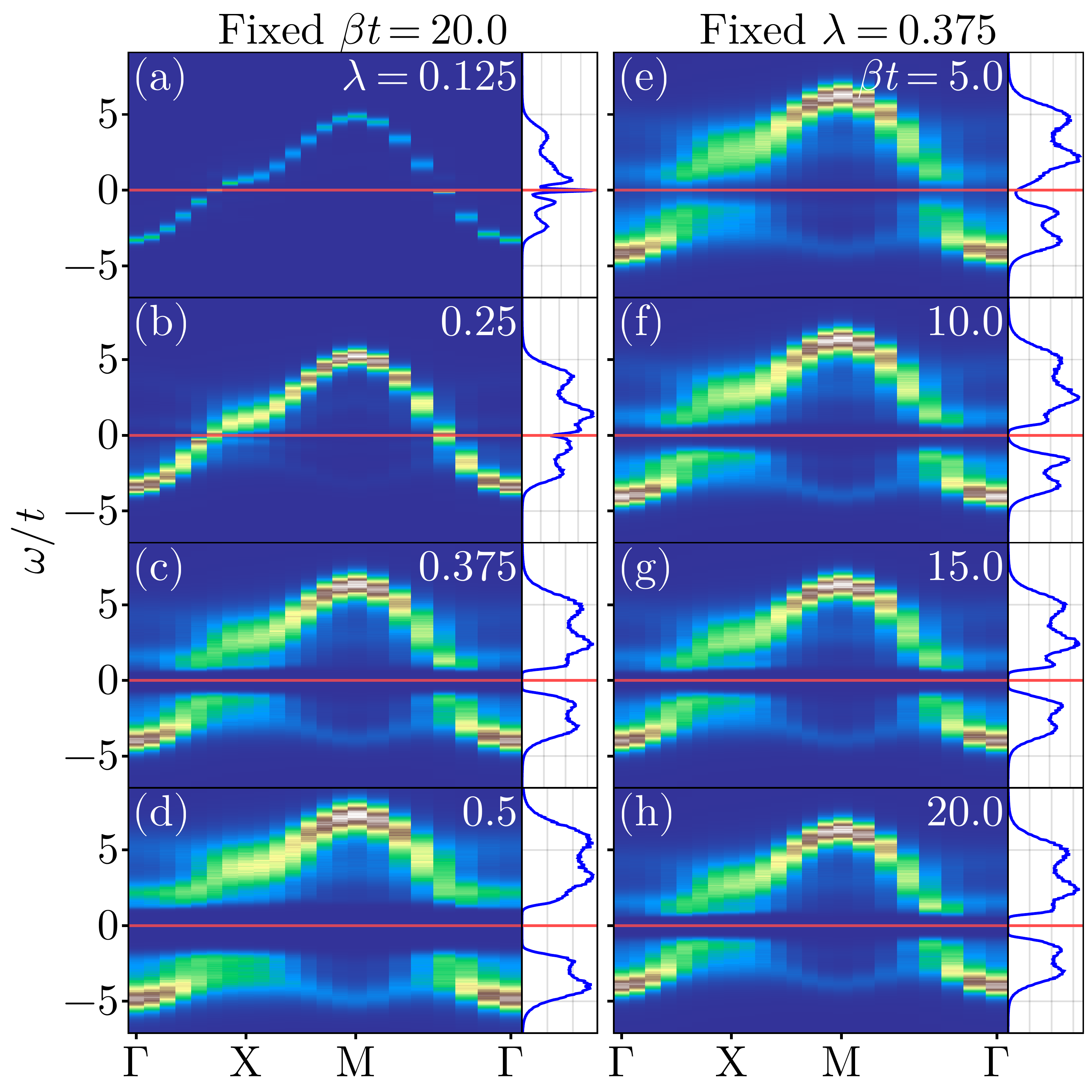}
    \caption{Single-particle spectral functions for Holstein model at $\langle n\rangle=0.7$ and $\Omega /t=0.5$. Panels (a)-(d) show results for fixed $\beta t=20$ and select $\lambda$ values, while panels (e)-(h) show results for $\lambda=0.375$ and select inverse temperatures $\beta$. The panels to the right of each spectral function show the density of states $N(\omega)$.}
    \label{fig:Spec0.7o0.5}
\end{figure}

Figure~\ref{fig:CondHalf}(b) shows the evolution of $\sigma(\omega)$ as the system is cooled for a fixed coupling $\lambda=0.25$. Note, $T_\mathrm{cdw} \approx t/5$ for these parameters~\cite{Weber2018TwoD}. At $\beta t=5$, $\sigma(\omega)$ has a finite dc conductivity and a dominant peak centered at $\omega/t \approx 1.4$. The spectra can be reasonably approximated by a sum of a Drude peak centered at $\omega = 0$ and an additional broad Gaussian peak at finite energy (not shown). These peaks reflect a weak residual metallic response despite the system being in close proximity to the transition temperature and the polaronic dressing of the carriers and transitions between the flattened portions of the spectra at the $X$ point, shown in Fig.~\ref{fig:SpecHalf}(e). Lowering the temperature to $\beta t\geq 10$ suppresses the Drude response completely, leaving only  transitions across the \gls*{CDW} gap. Lowering the temperature further produces no notable changes in the spectra, suggesting that $\beta t= 10$ is sufficient to access the ground-state properties of the system for this cluster size. 

\subsection{Intermediate doping, $\langle n \rangle = 0.7$}
\begin{figure}[t]
    \centering
    \includegraphics[width=\linewidth]{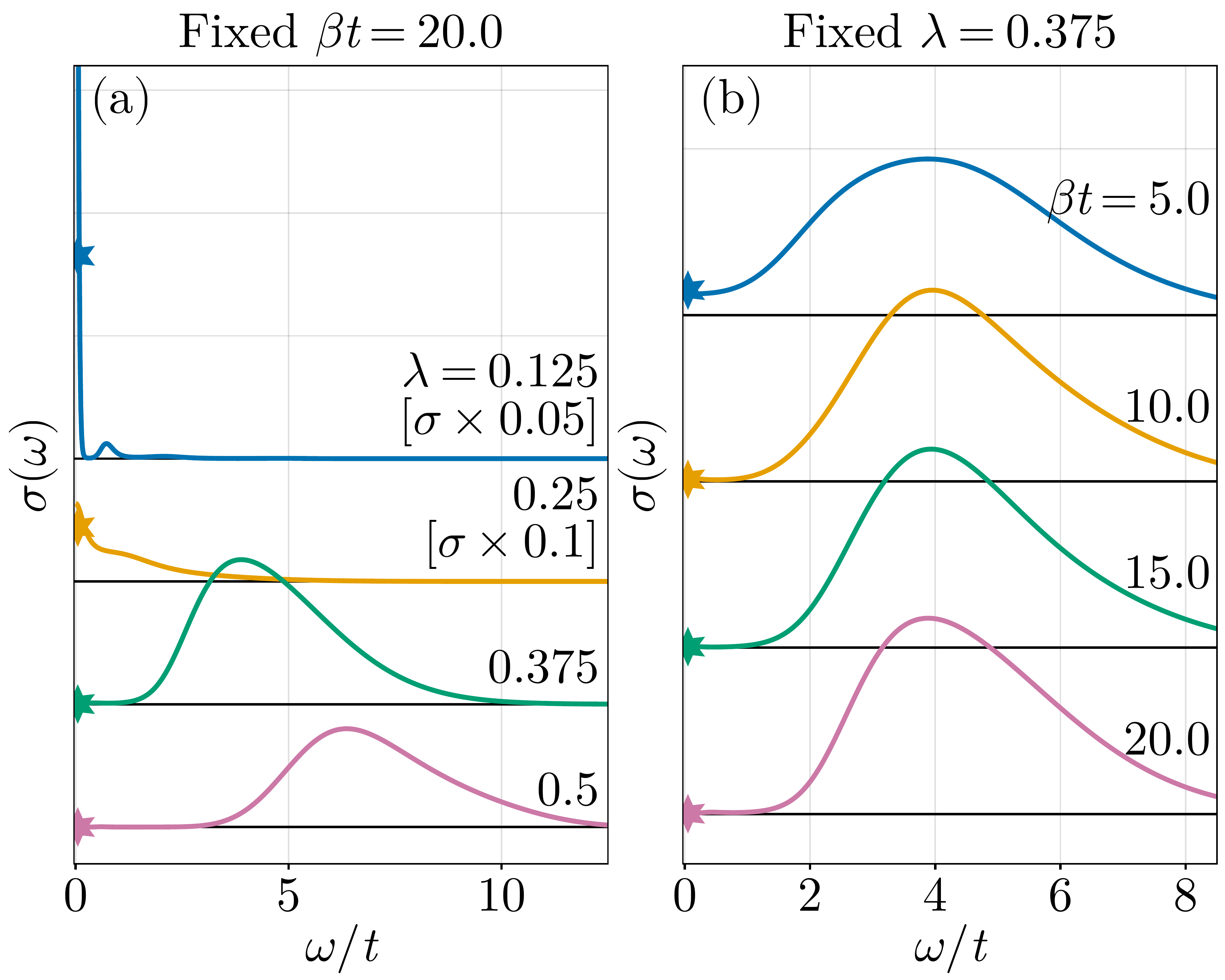}
    \caption{Optical conductivity for the doped Holstein model with $\langle n\rangle=0.7$ and $\Omega/t=0.5$. The star markers at $\omega/t=0$ are the value of dc conductivity obtained using the imaginary axis proxy $\sigma_\mathrm{dc}$ given by Eq.~\eqref{eq:sigma_dc}. Panel (a) shows results for fixed inverse temperatures $\beta t= 20$ and select $\lambda$ while panel (b) shows results for fixed $\lambda = 0.375$ and select $\beta$.  }
    \label{fig:Cond0.7o0.5}
\end{figure}

We now turn our attention to a doped case with an average filling of $\langle n\rangle=0.7$ and a phonon energy of $\Omega/t=0.5$. Fig.~\ref{fig:Spec0.7o0.5} shows the spectral function in a format similar to Fig.~\ref{fig:SpecHalf}. As with the half-filled case, the low-temperature spectra at $\lambda = 0.125$ [Fig.~\ref{fig:Spec0.7o0.5}(a)] closely resemble the noninteracting band. Increasing the coupling to $\lambda=0.25$ produces results consistent with Migdal theory~\cite{Nosarzewski2021spectral}; the dispersion acquires a kink-like structure at $\omega = \pm \Omega$ with increased broadening at higher energies. The density of states also shows a slight depletion near the Fermi level, resembling a pseudogap as reported previously~\cite {Esterlis2019pseudogap}. Increasing the coupling further results in a fully gapped electronic structure [Fig.~\ref{fig:Spec0.7o0.5}(c) \& (d)]. For $\lambda=0.375$, we also begin to observe the formation of a Franck-Condon-like shake-off state at higher binding energy, which is most evident near the $\Gamma$-point. As we will argue below, the gap forming at the Fermi level does not necessarily indicate the formation of an insulating state; rather, it reflects the binding energy associated with bipolaron formation~\cite{Kovac2025Signature}.  It is also worth noting that we can resolve only a single replica band rather than a series of shake-off states up to the strongest coupling value.\footnote{We have observed weak indications of a two-phonon feature on larger clusters, see App.~\ref{AppendixFinite}, which suggests finite size effects may play a small role here in obscuring the first higher order features; however, even on the larger lattices, we do not resolve the ``ladder'' of shake-off states that are expected given the large binding energy of the bipolaron.}
Similar observations---suppressed spectral weight with bipolaron  formation and weak higher-order replica features---were reported in recent \gls*{DMRG} study of the \gls*{1D} Hubbard-Holstein model~\cite{Kovac2025Signature}. 

Figures~\ref{fig:Spec0.7o0.5}(e)-(h) show the evolution of the spectra for a fixed $\lambda=0.375$ as a function of inverse temperature. The results have indications of a gap opening at the Fermi level at high temperature ($\beta t=5$), which becomes fully formed by $\beta t=10$. At all temperatures, we also resolve backfolded spectral weight and a single Franck-Condon shake-off state, with the spectral features becoming sharper as the temperature is lowered. 

\begin{figure}[t]
    \centering
    \includegraphics[width=\linewidth]{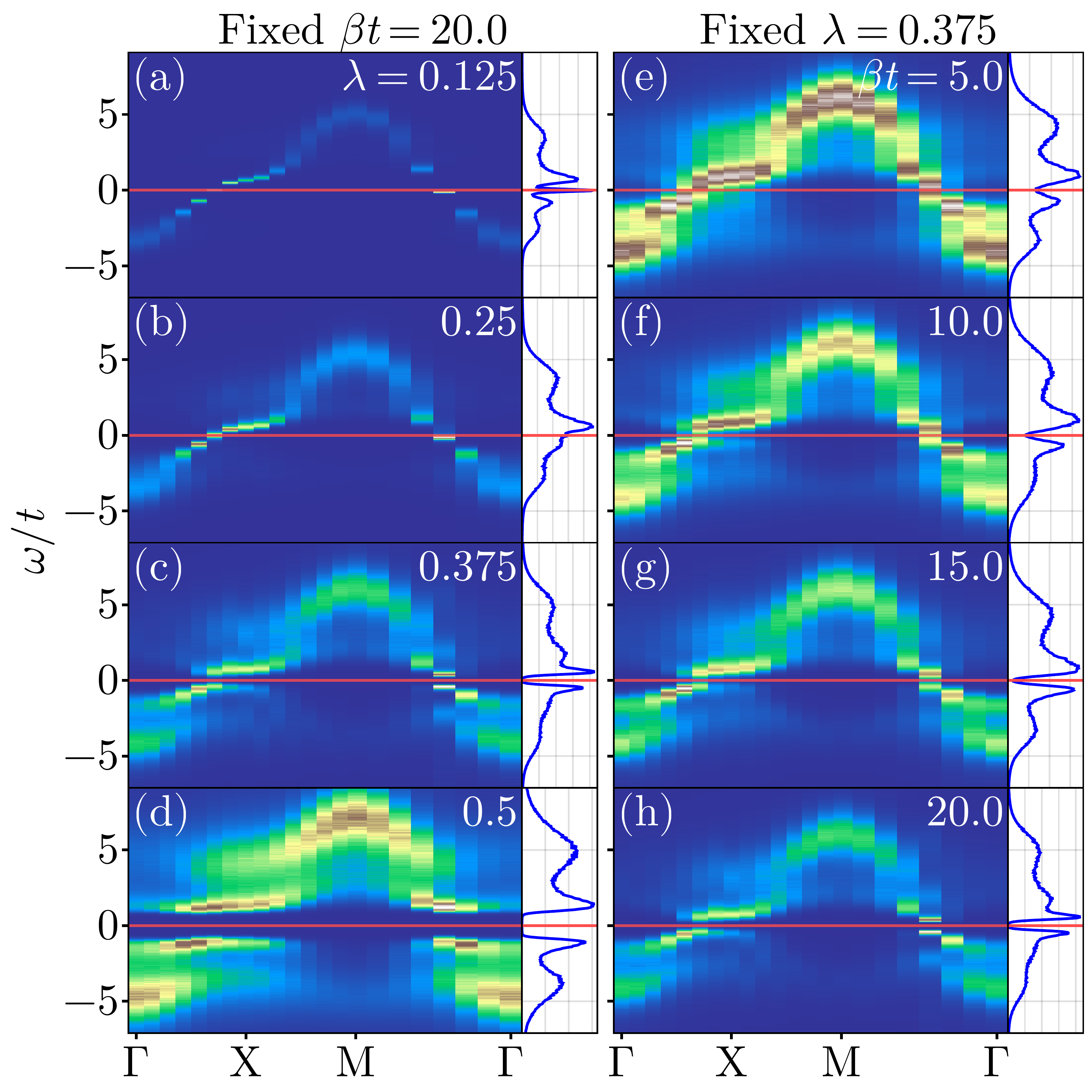}
    \caption{Single particle spectral functions for $\langle n\rangle=0.7$ and $\Omega/t=2$. (a)-(d) $\beta t=20$ spectral functions for select $\lambda$ values, as indicated. (e)-(h) $\lambda=0.375$ spectral functions for select $\beta t$, as indicated. The panels to the right of each spectral function show the density of states $N(\omega)$. }
    \label{fig:Spec0.7o2.0}
\end{figure}

Figure~\ref{fig:Cond0.7o0.5} examines the optical conductivity for the same parameters used in Fig.~\ref{fig:Spec0.7o0.5}. Fig.~\ref{fig:Cond0.7o0.5}(a) shows low-temperature ($\beta t=20$) results as a function of $\lambda$, where each curve has been offset for clarity. For the lowest coupling strength $\lambda=0.125$, the spectrum is dominated by a sharp Drude peak indicative of metallic behavior. As the coupling is increased to $\lambda=0.25$, the Drude response broadens while spectral weight shifts into a peak at $\omega/t\approx0.8$. 
Increasing $\lambda$ further fully suppresses the Drude peak and shifts the location of secondary peak to higher energies.  This behavior closely resembles the evolution of the optical conductivity calculated in the dilute limit using the momentum-average and diagrammatic \gls*{QMC} approaches~\cite{Goodvin2011optical} but is obtained here for a finite carrier concentration. The evolution of $\sigma(\omega)$ shown in Fig.~\ref{fig:Cond0.7o0.5}(a) thus evidences a low-temperature metal-to-insulator crossover with increasing $\lambda$.

Figure~\ref{fig:Cond0.7o0.5}(b) plots the temperature evolution of $\sigma(\omega)$ for a fixed $\lambda=0.375$. Recall $A(\boldsymbol{k},\omega)$ for this value of the coupling exhibited a clear spectral gap for $\beta t\ge 5$. The optical conductivity, in this case, is dominated by a broad peak at finite energy, consistent with the strong-coupling results shown in panel (a). However, we also obtain a small dc conductivity at high temperatures $\beta t=5$, which can be inferred from both the analytically continued spectra and the imaginary time axis proxy for $\sigma(0)$ [blue $\star$, see also Eq.~\eqref{eq:sigma_dc}].

\begin{figure}[t]
    \centering
    \includegraphics[width=\linewidth]{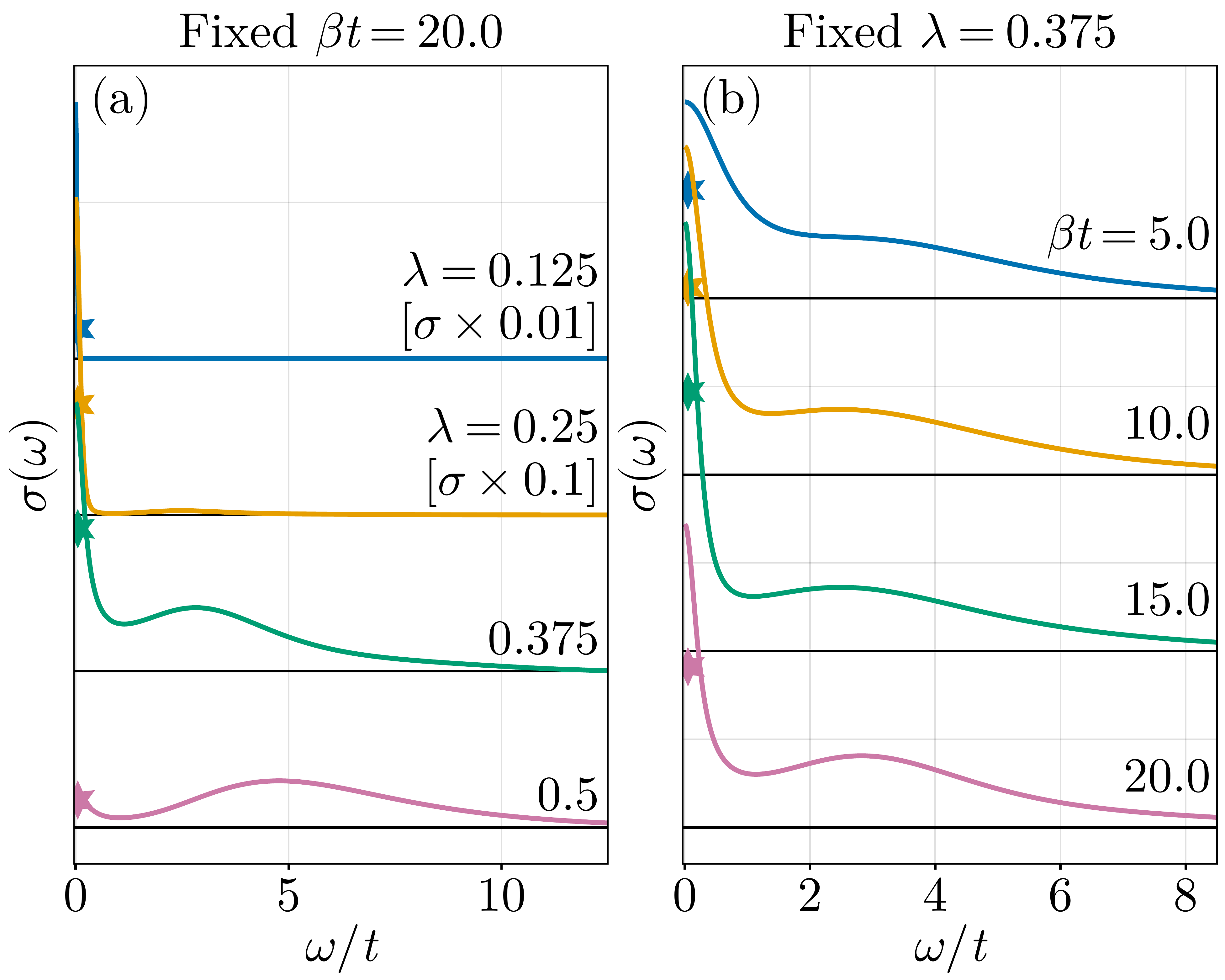}
    \caption{Optical conductivity for the doped Holstein model with $\langle n\rangle=0.7$ and $\Omega/t=2$. The star markers at $\omega/t=0$ are the value of dc conductivity obtained using the imaginary axis proxy $\sigma_\mathrm{dc}$ given by Eq.~\eqref{eq:sigma_dc}. Panel (a) shows results for fixed inverse temperatures $\beta t= 20$ and select $\lambda$ while panel (b) shows results for fixed $\lambda = 0.375$ and select $\beta$. }
    \label{fig:Cond0.7o2.0}
\end{figure}

From Fig.~\ref{fig:Spec0.7o0.5}, we interpret the spectral gap in the single-particle properties seen at lower temperatures and higher couplings as being indicative of bipolaron formation. In this case, the gap reflects the energy needed to break the bound electron pair before one can be removed~\cite{Kovac2025Signature}. The dc conductivity in \ref{fig:Cond0.7o0.5} then suggests that the bipolarons have a small amount of conductivity at high temperatures (albeit with a significantly reduced mobility) and become trapped on the time scales of our \gls*{DQMC} simulations at low temperatures. We will provide further evidence for this interpretation of the data in Fig.~\ref{fig:condvsDOS} but also note that it is consistent with previous \gls*{DQMC} studies that concluded that Migdal's theorem breaks down in the Holstein model for $\lambda \gtrapprox 0.4$ for $\Omega/t \sim O(1)$~\cite{Esterlis2018breakdown, Nosarzewski2021superconductivity}.

According to Migdal's theorem, vertex diagrams are weighted by a prefactor $\lambda\times(\Omega / E_\mathrm{F})$. Thus, one naively expects that polaronic effects will become more pronounced in the antiadiabatic limit $\Omega/t \rightarrow \infty$ for a fixed $\lambda$. At the same time, the tendency for the system to form large lattice distortions will be suppressed by the stiffer spring constants $K = M\Omega^2$ associated with the harmonic lattice potential as $\Omega$ is increased, which could increase polaron mobility. Motivated by these expectations, Figs. \ref{fig:Spec0.7o2.0} and \ref{fig:Cond0.7o2.0} examine $A(\boldsymbol{k},\omega)$ and $\sigma(\omega)$, respectively, for $\langle n\rangle=0.7$ but an increased $\Omega/t=2$, following a similar format as the previous cases. 

At low temperature and weak coupling ($\lambda=0.125$, $0.25$), the spectral function and density of states again resemble expectations for a weakly dressed system; spectral weight broadens for $|\omega| > \Omega$ and a single kink-like band renormalization appears at the phonon energy. The spectra begin to gap for $\lambda\geq 0.375$, at which point the kink-like renormalization shifts to $\pm |\Omega+\Delta|$, where $\Delta$ is the size of the gap in the single-particle spectrum, and more closely resemble the typical two-pole solution obtained in Migdal theory~\cite{Engelsberg1963coupled}. For the strongest coupling, $\lambda=0.5$, we further find that the sharper quasi-particle portion of the spectra flattens significantly, as does the portion of the band at higher binding energies. As with the lower phonon energy, here we do not resolve the formation of additional replicas typically associated with Franck-Condon shake-off states. 

\begin{figure}[t]
    \centering
    \includegraphics[width=\linewidth]{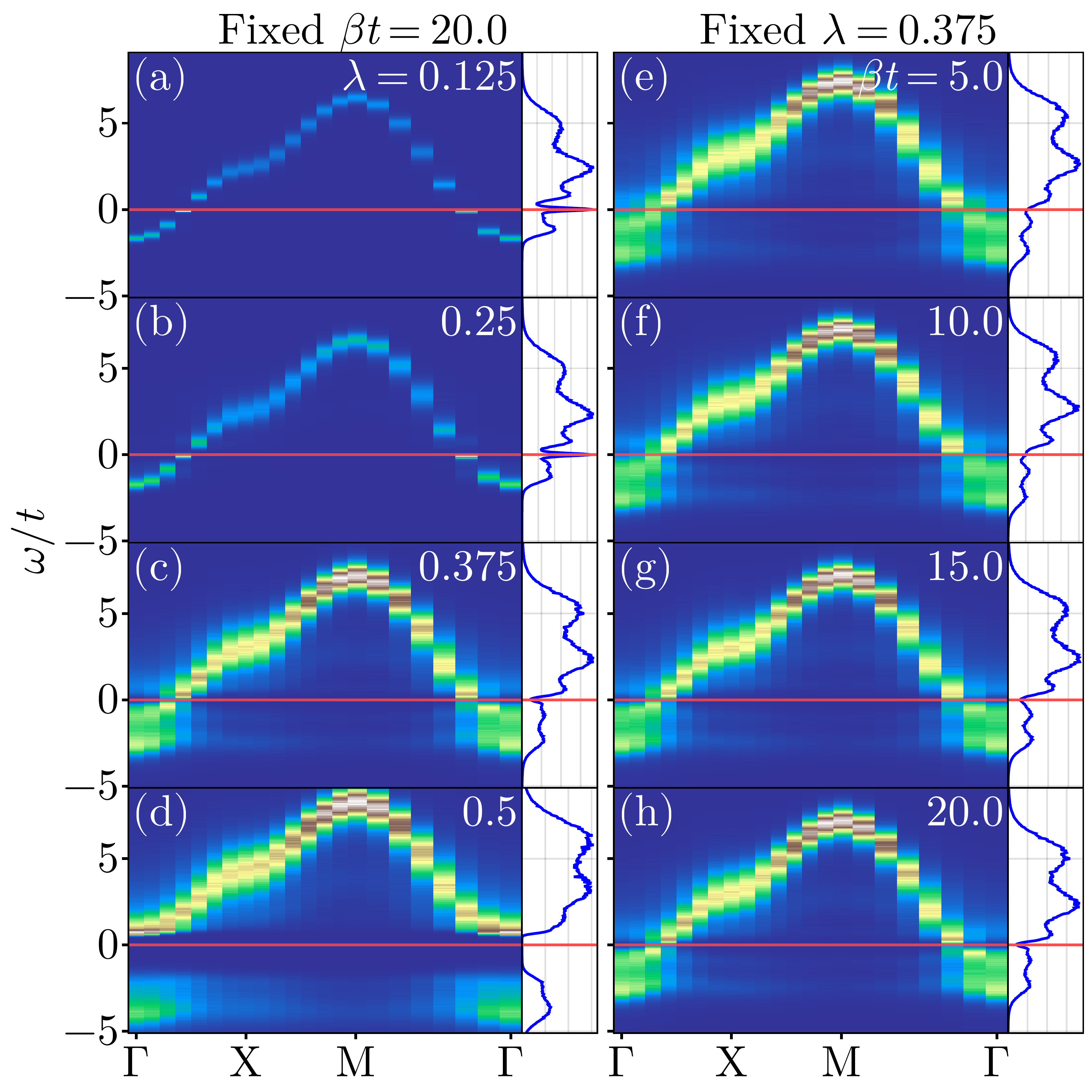}
    \caption{Single-particle spectral functions for $\langle n\rangle=0.3$ and $\Omega/t=0.5$. (a)-(d) $\beta t=20$ spectral functions for select $\lambda$ values, as indicated. (e)-(h) $\lambda=0.375$ spectral functions for select $\beta t$, as indicated. The panels to the right of each spectral function show the density of states $N(\omega)$.}
    \label{fig:Spec0.3o0.5}
\end{figure}

\begin{figure}[t]
    \centering
    \includegraphics[width=\linewidth]{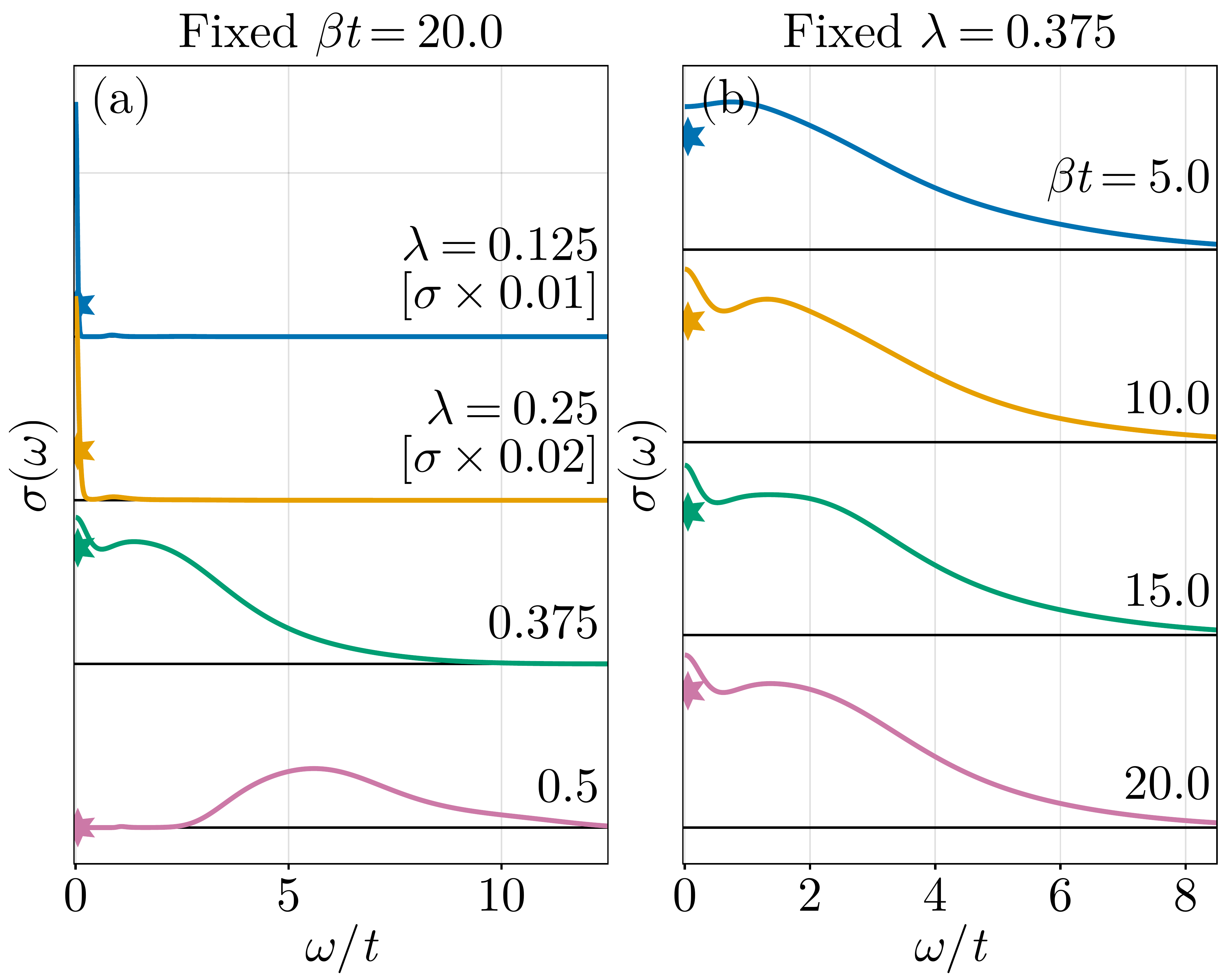}
    \caption{Optical conductivity for the doped Holstein model with $\langle n\rangle=0.3$ and $\Omega/t=0.5$. The star markers at $\omega/t=0$ are the value of dc conductivity obtained using the imaginary axis proxy $\sigma_\mathrm{dc}$ given by Eq.~\eqref{eq:sigma_dc}. Panel (a) shows results for fixed inverse temperatures $\beta t= 20$ and select $\lambda$ while panel (b) shows results for fixed $\lambda = 0.375$ and select $\beta$. }
    \label{fig:Cond0.3o0.5}
\end{figure}

The optical conductivity, shown in Fig.~\ref{fig:Cond0.7o2.0},  also exhibits interesting behavior for this value of $\Omega$. For weak coupling, the spectrum has a Drude response that broadens as the coupling increases. However, unlike for $\Omega/t = 0.5$, we find that the dc response inferred from both $\sigma(\omega)$ and the imaginary time proxy is nonzero at all couplings. This behavior again suggests that the system hosts mobile bipolarons at low temperature. Two additional observations can be made at this point: first, the agreement between the proxy dc conductivity and the value inferred from the analytically continued optical conductivity does not agree as well as it did in the $\Omega/t = 0.5$ case shown in Fig.~\ref{fig:Cond0.7o0.5}. This discrepancy is likely due to the peaked structure of kernel in 
Eq.~\eqref{eq:ACboson} at $\tau=\beta/2$, where it becomes $2\cosh{\omega\beta/2}$. This broadened function typically yields proxy measurements with underreported values for large Drude peaks and slightly larger values in cases where there is no dc conductivity but there is a secondary peak at low energies near $\omega/t=0$. Second, for the $\Omega/t = 0.5$, the backfolding feature is much clearer than for $\Omega/t = 2$ at this filling. This difference suggests that the $\Omega/t = 2$ case does not exhibit the same degree of short-range $\boldsymbol{Q}= (\pi,\pi)$ order compared to the $\Omega/t = 0.5$ case. Combined, the results suggest that immobile bipolarons phase separate into domains of \gls*{CDW} order at $\Omega/t = 0.5$ but remain mobile at $\Omega/t= 2$. 

\subsection{Dilute fillings, $\langle n \rangle = 0.3$}
\begin{figure}[t]
    \centering
    \includegraphics[width=\linewidth]{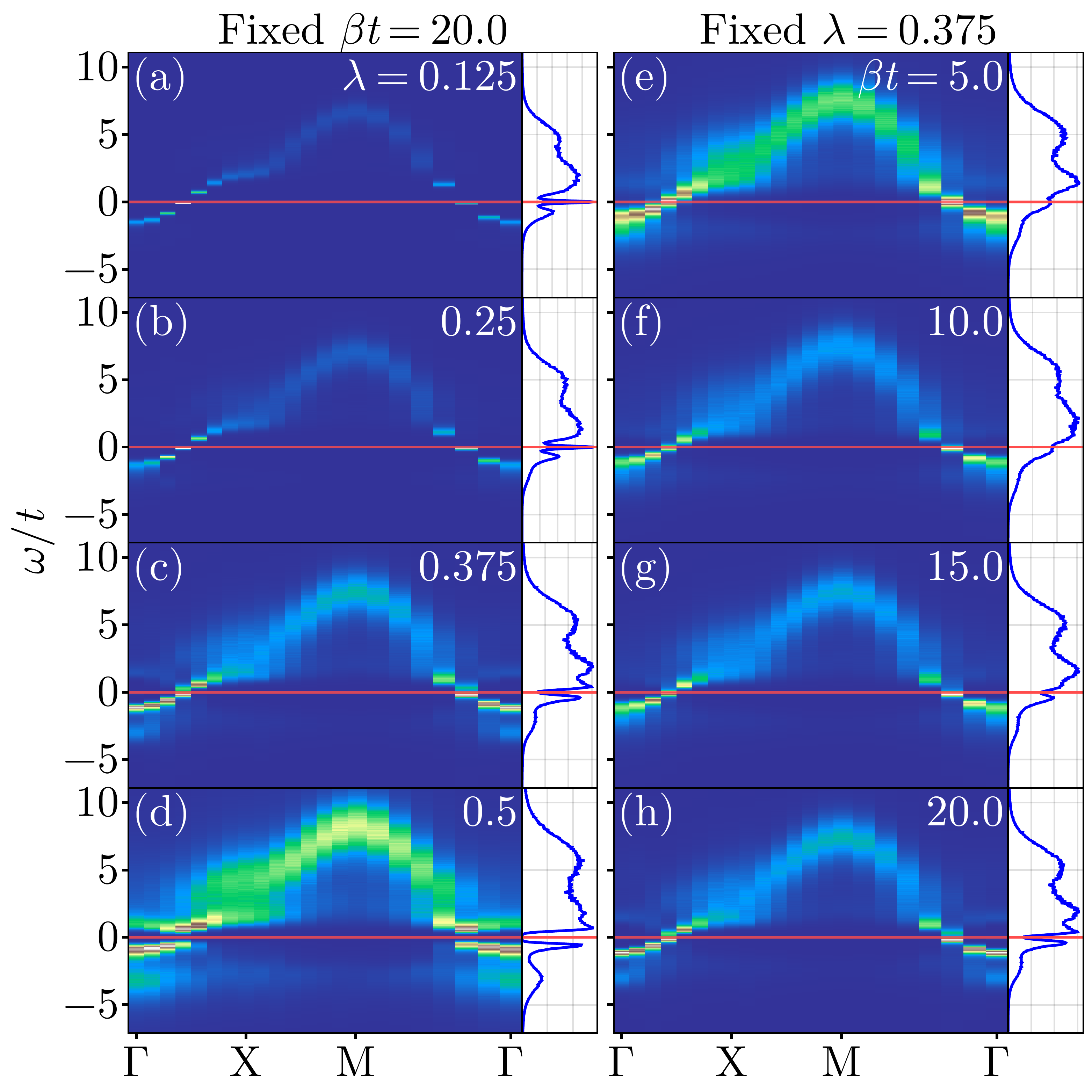}
    \caption{Single-particle spectral functions for $\langle n\rangle=0.3$ and $\Omega/t=2$. (a)-(d) $\beta t=20$ spectral functions for select $\lambda$ values, as indicated. (e)-(h) $\lambda=0.375$ spectral functions for select $\beta t$, as indicated. The panels to the right of each spectral function show the density of states $N(\omega)$.}
    \label{fig:Spec0.3o2.0}
\end{figure}

\begin{figure}[t]
    \centering
    \includegraphics[width=\linewidth]{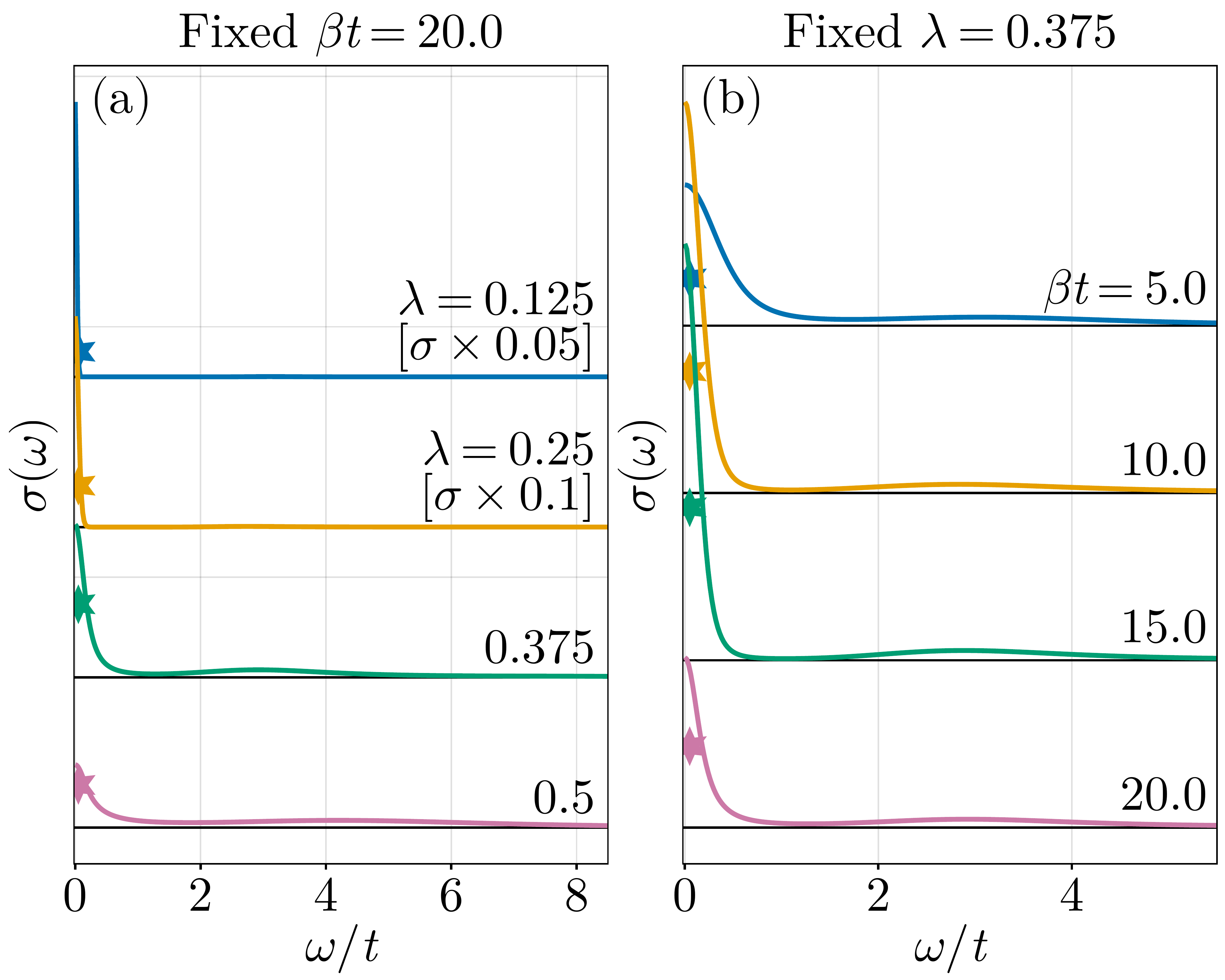}
    \caption{Optical conductivity for the doped Holstein model with $\langle n\rangle=0.3$ and $\Omega/t=2$. The star markers at $\omega/t=0$ are the value of dc conductivity obtained using the imaginary axis proxy $\sigma_\mathrm{dc}$ given by Eq.~\eqref{eq:sigma_dc}. Panel (a) shows results for fixed inverse temperature $\beta t = 20$ and select $\lambda$, while panel (b) shows results for fixed $\lambda = 0.375$ and select $\beta$.}
    \label{fig:Cond0.3o2.0}
\end{figure}

We now turn our attention to a dilute case, $\langle n \rangle = 0.3$. Figs.~\ref{fig:Spec0.3o0.5} and \ref{fig:Spec0.3o2.0} plot $A(\boldsymbol{k},\omega)$ at this filling for $\Omega/t = 0.5$ and $2$, respectively, while Figs.~\ref{fig:Cond0.3o0.5} and \ref{fig:Cond0.3o2.0} show the corresponding optical conductivities. For this value of the filling, the noninteracting band minimum is $\epsilon(\boldsymbol{k} = 0)= -1.66t$, which places it slightly below (above) the phonon energy in Fig.~\ref{fig:Spec0.3o0.5} (Fig.~\ref{fig:Spec0.3o2.0}). 

As with the previous filling values, the spectra for the $\lambda=0.125$ and $0.25$ cases 
are only slightly dressed by the \gls*{eph} interaction. The spectral broadening is more noticeable for the higher coupling, $\lambda=0.25$, along the entire high-symmetry path, while the kink structure is difficult to discern due to the momentum resolution and the low filling. The \gls*{DOS} for both couplings are nearly identical with a peak near the Fermi energy. The band renormalizations become more apparent for a coupling strength of $\lambda=0.375$ [Figs.~\ref{fig:Spec0.3o0.5}(c) and \ref{fig:Spec0.3o2.0}(c)]; for $\Omega/t=0.5$, we can resolve a clear kink feature, and the formation of a phonon replica state appears below the band minimum near the $\Gamma$ point. 
A pronounced broadening is also apparent throughout the spectra and the \gls*{DOS} trends towards a pseudogap structure with depleted weight at the Fermi level. Increasing the coupling to $\lambda=0.5$ [Figs.~\ref{fig:Spec0.3o0.5}(d) and ~\ref{fig:Spec0.3o2.0}(d)] results in a full gap at $E_\mathrm{F}$, again indicating the bipolaron formation. The spectra are further broadened, and the kink-like structure in the unoccupied states becomes more apparent. Figs.~\ref{fig:Spec0.3o0.5}(e)-(h) and Figs.~\ref{fig:Spec0.3o2.0}(e)-(h) show the evolution of the spectra as a function of temperature, again for $\lambda = 0.375$. Across all temperatures we see a very weak evolution of the spectral functions and densities of states, with the pseudogap depletion occurring as temperature decreases.

These trends are also evident in the optical conductivity plots shown in Figs. ~\ref{fig:Cond0.3o0.5} and \ref{fig:Cond0.3o2.0}. 
Interestingly, the spectrum for $\Omega/t = 0.5$ is fully gapped for the largest coupling $\lambda = 0.5$, indicating insulating behavior. Conversely, for $\Omega/t = 2$, we obtain a finite conductivity at this coupling, suggesting that the bipolarons remain mobile for this value of the phonon energy. 

\subsection{Crossover diagrams}

\begin{figure}[t]
    \centering
    \includegraphics[width=\linewidth]{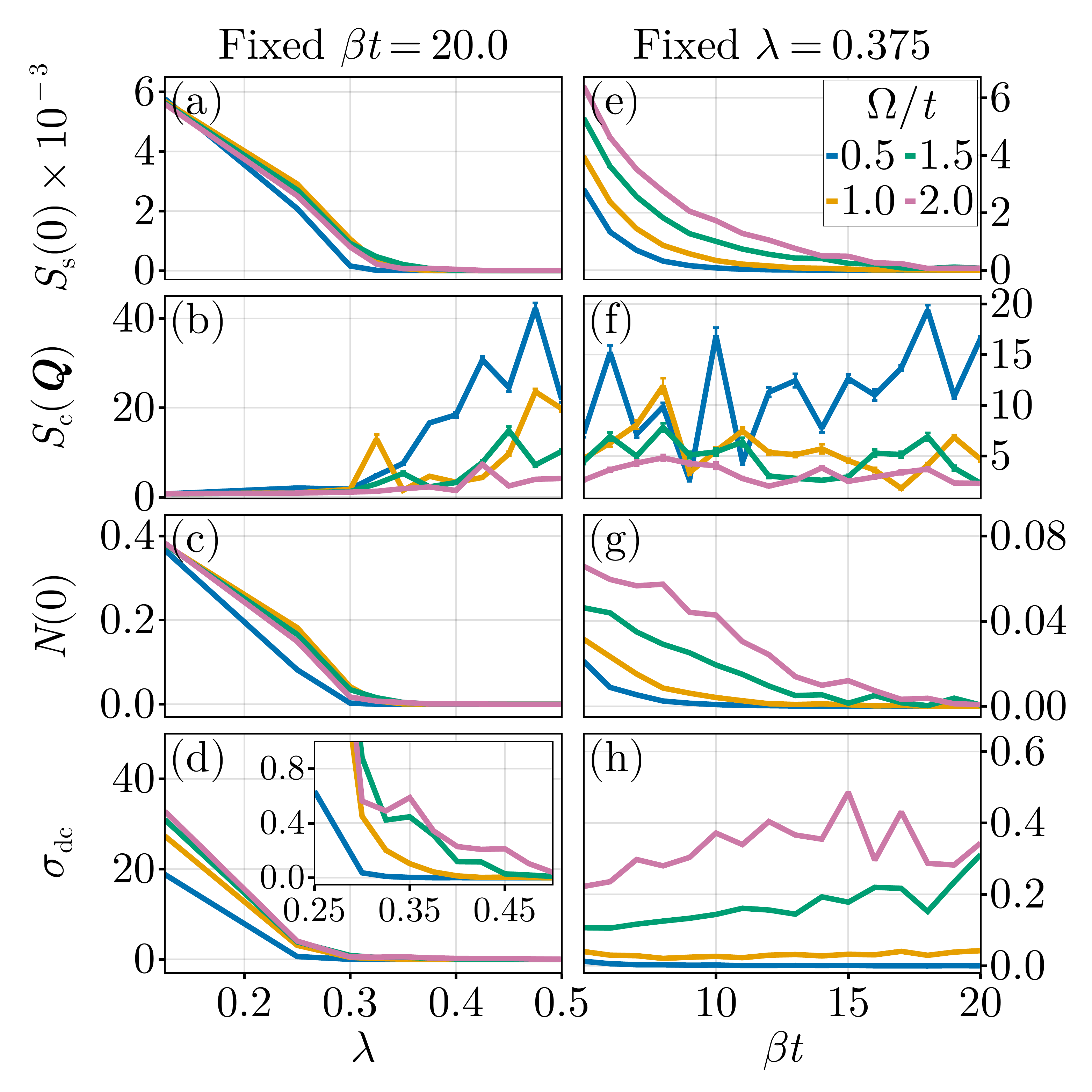}
    \caption{Assorted equal time measurements for the doped Holstein model with $\langle n\rangle=0.7$ at various values of $\Omega$. Results are shown for (a) the uniform spin structure factor, (b) $\boldsymbol{Q}=(\pi,\pi)$ charge structure factor, (c) the DOS at the Fermi energy, (d) and dc conductivity as a function of $\lambda$ and fixed $\beta t=20$. Panels (e)-(h) plot similar data as a function of inverse temperature $\beta$ and fixed $\lambda=0.375$.}
    \label{fig:condvsDOS}
\end{figure}

To place our interpretation of the dynamical correlation functions on a firmer footing, we now examine the evolution of the several other quantities as a function of inverse temperature, \gls*{eph} coupling, and phonon energy (i.e., retardation effects). To this end, Fig.~\ref{fig:condvsDOS} plots several observables for $\langle n\rangle=0.7$ and $\Omega /t = 0.5$, $1$, $1.5$, and $2$. These include the uniform spin structure factor $S_\mathrm{s}(\boldsymbol{q} =0)$, the charge structure factor at the \gls*{CDW} wave vector $S_\mathrm{c}(\boldsymbol{Q})$, the density of states at the Fermi level $N(0)$, and dc conductivity, both obtained through analytic continuation. (We obtain qualitatively similar results if we use the imaginary time proxies for these quantities.) 

Figures~\ref{fig:condvsDOS}(a) and \ref{fig:condvsDOS}(e) show that the uniform spin structure factor $S_\mathrm{s}(\boldsymbol{q}=0)$ approaches zero for both increasing coupling strength and decreasing temperature. This behavior is observed for all phonon energies, but the suppression is more rapid for smaller values of $\Omega$. As the spin structure factor is suppressed, we observe a corresponding increase in the charge structure factor $S_\mathrm{c}(\boldsymbol{Q})$ [Figs.~\ref{fig:condvsDOS}(b) and \ref{fig:condvsDOS}(f)], and a suppression in the density of states at the Fermi level [Figs.~\ref{fig:condvsDOS}(c) and Fig.~\ref{fig:condvsDOS}(g)]. This behavior is broadly consistent with bipolaron singlet formation at large $\lambda$ and low temperatures. 

A picture of bipolaron formation is further supported by the behavior in the dc conductivity (see below) and large fluctuations in charge structure factor data. In general, we find $S_\mathrm{c}(\boldsymbol{Q})$ exhibits large fluctuations that are much greater than the error inferred from an analysis of the Markov chains themselves (as indicated by the error bars), particularly for large $\lambda$ and small $\Omega/t$. This behavior can be traced to a tendency for the bipolarons in the system to clump together during a simulation forming regions of short-range \gls*{CDW} order, consistent with previous reports~\cite{Nosarzewski2021superconductivity}. We have explicitly confirmed this behavior by examining the distribution of filling and displacement measurements in our binned measurements (not shown). 

\begin{figure}[t]
    \centering
    \vspace{0.5cm}
    \includegraphics[width=0.7\linewidth]{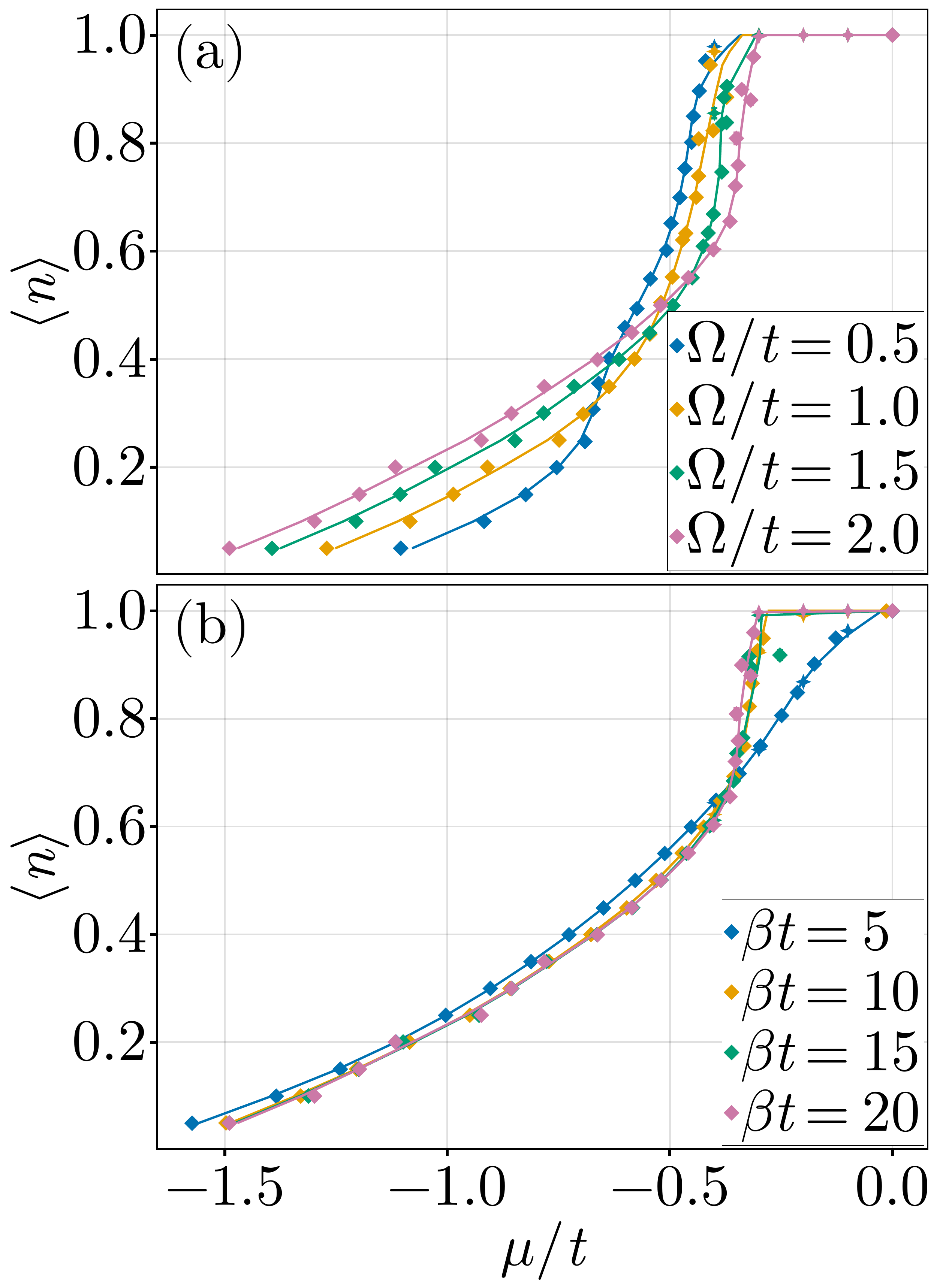}
    \caption{(a) The average filling $\langle n\rangle$  vs. chemical potential $\mu/t$ for $\lambda=0.375$ and $\beta t=20$ for all values of $\Omega$. Star points represent simulations run at fixed $\mu/t$. (b) $\langle n\rangle$ vs. $\mu/t$ for fixed $\lambda=0.375$ and $\Omega/t=2$ for selected values of $\beta$.}
    \label{fig:Mu_vs_N}
\end{figure}

Indications of short-range phase separation also appear in the behavior of the average filling $\langle n \rangle$ as a function of $\mu$. For example, Fig.~\ref{fig:Mu_vs_N}(a) shows that average filling at low temperatures exhibits a rapid, nearly vertical upturn as chemical potential is tuned toward $\mu/t\approx -0.5$ for all phonon frequencies. This behavior signals a sharp increase in the electronic compressibility $\kappa =\frac{1}{\langle n \rangle^2} \left(\partial\langle n \rangle /\partial \mu\right)_T$, which is indicative of large charge fluctuations between coexisting regions and is often observed near first order phase transitions. This behavior disappears upon warming, as shown in Fig.~\ref{fig:Mu_vs_N}(b) for $\Omega/t=2$ and $\lambda = 0.375$.  

Turning to the conductive properties, Fig.~\ref{fig:condvsDOS}(d) plots the dc conductivity at $\beta=20/t$ as a function of $\lambda$. At $\lambda\leq0.25$, the conductivity is finite for all values of $\Omega$, indicating that the system hosts mobile charge carriers. However, it drops to zero for $\lambda \gtrapprox 0.4$ and is suppressed more rapidly for small phonon energies, as highlighted in the inset. Increasing the phonon frequency toward the antiadiabatic regime, e.g., $\Omega/t = 1.5$, $2$, extends the region of small finite conductivity to even higher values of the coupling $\lambda=0.5$. This behavior is consistent with a transition from polaronic quasiparticles at small coupling to heavy (bi)polaronic carriers at strong coupling, which have a strong tendency toward self-trapping. This view is also supported by the temperature dependence of the conductivity, which is shown in Fig.~\ref{fig:condvsDOS}(h) for $\lambda=0.375$. In this case, the dc conductivity is small and mostly temperature-independent at low phonon energies, suggesting insulating behavior, but it differs notably at higher phonon energies. For $\Omega = 1.5t$, $\sigma_\mathrm{dc}$ decreases with increasing temperature (decreasing $\beta$), indicative of metallic transport and Bloch-like polaronic states.  Conversely, for $\Omega = 2t$, $\sigma_\mathrm{dc}$ has a nonmonotonic behavior, initially increasing as the system is cooled from $\beta t = 5$ before ultimately decreasing at lower temperatures. This nonmonotonic behavior is highly suggestive of small (bi)polaron conduction where transport is governed by rare hopping events at low temperature, which gives way to thermal-assisted hopping as $T$ increases. Conversely, at high temperatures, thermal excitation of the lattice induces quasi-particle scattering and ultimately reduces the conductivity. 

\begin{figure}[t]
    \centering
    \includegraphics[width=\linewidth]{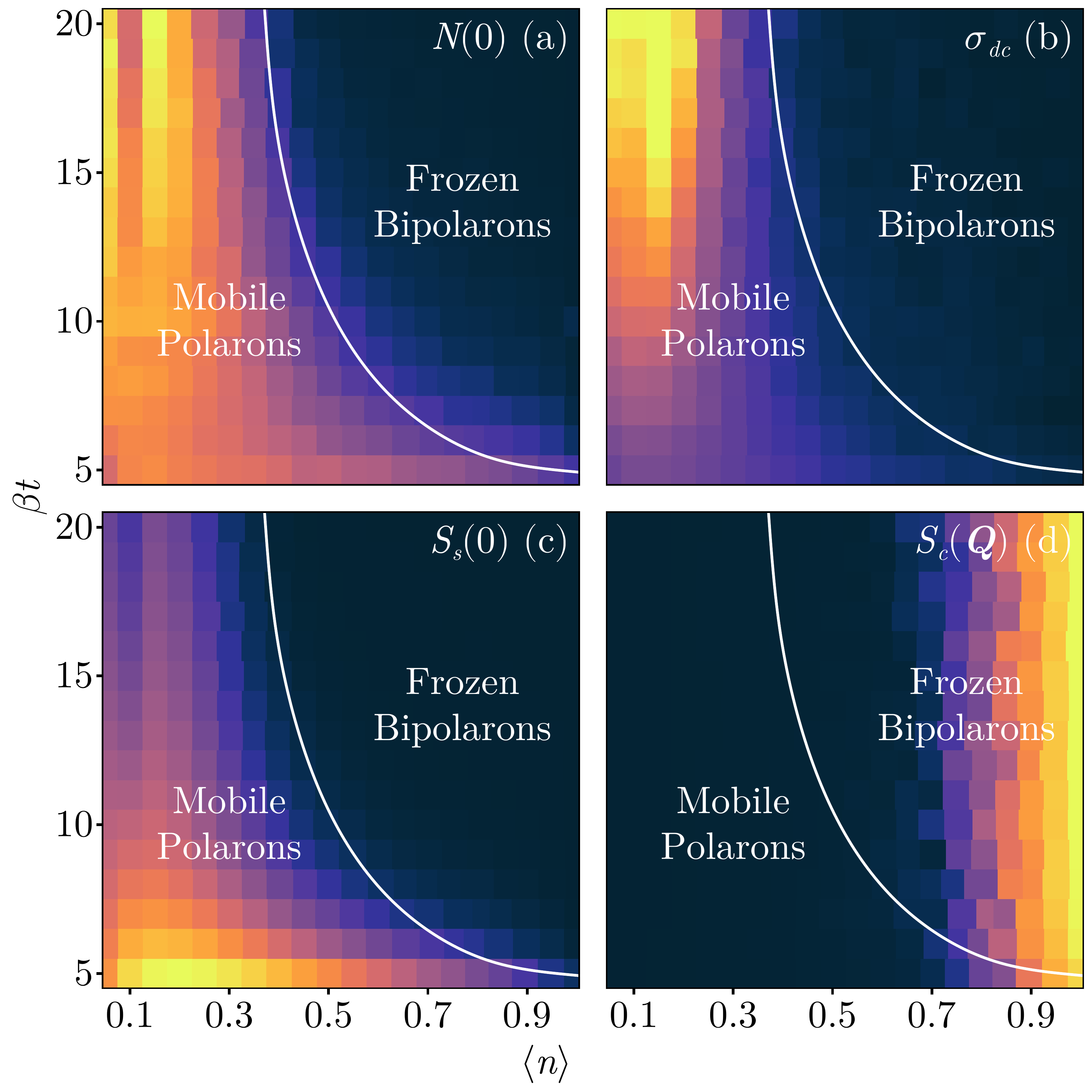}
    \caption{A crossover $\langle  n\rangle -T$ diagram highlighting different behaviors in the doped Holstein model for fixed $\lambda = 0.375$ and $\Omega/t=0.5$. Results are shown for 
    (a) the single-particle density of states at the Fermi energy $N(0)$ on a square root color scale, (b) the dc conductivity $\sigma_\mathrm{dc}(0)$ on a square root color scale [see Eq.~\eqref{eq:sigma_dc}], (c) the uniform spin structure factor $S_\mathrm{s}(\boldsymbol{q} = 0)$, and (d) charge structure factor $S_\mathrm{c}(\boldsymbol{Q}$) at the CDW wave vector $\boldsymbol{Q}=(\pi,\pi)$. The solid line indicates the contour along which $N(0) = 0.005$.}
    \label{fig:crossover_05}
\end{figure}

\begin{figure}[t]
    \centering
    \includegraphics[width=\linewidth]{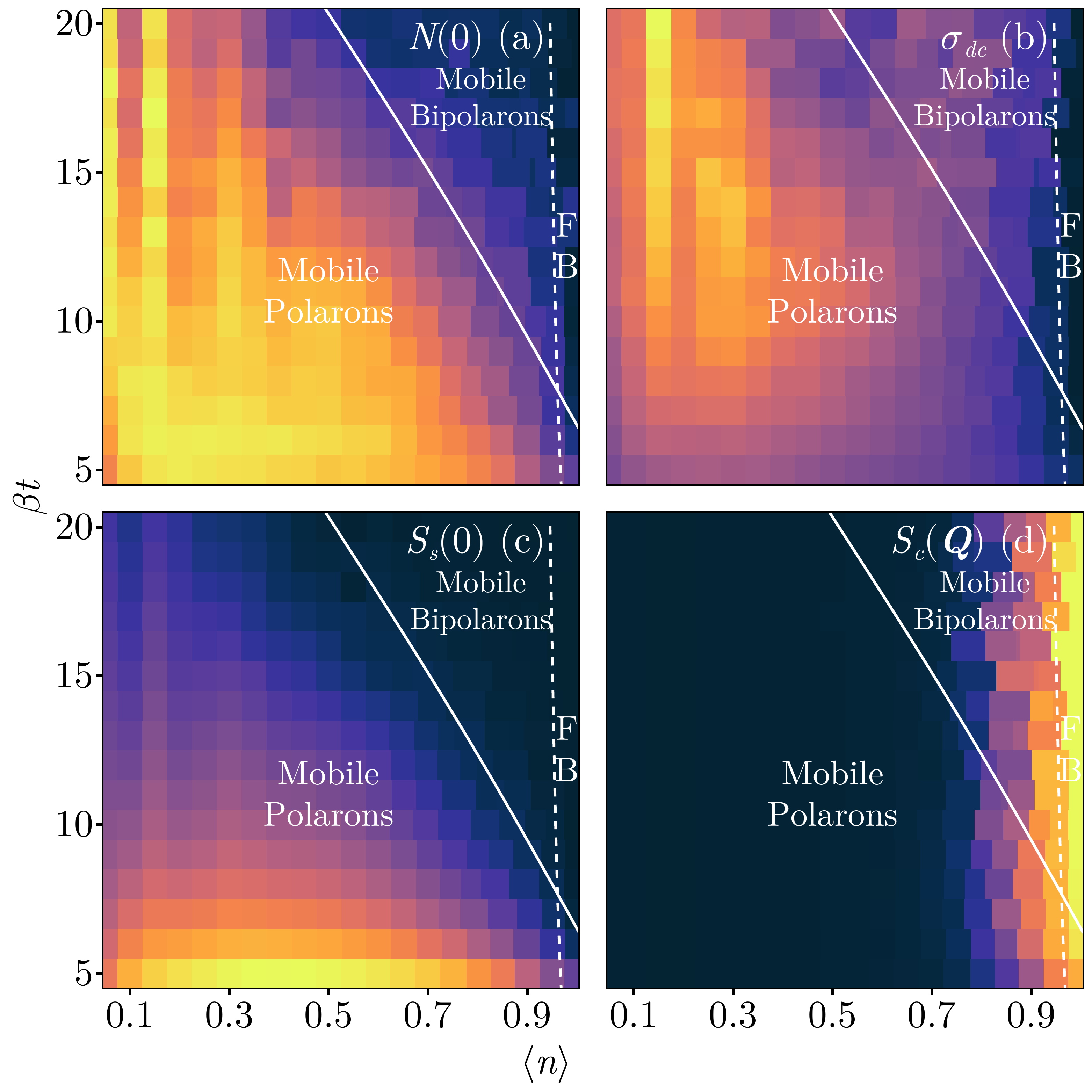}
    \caption{A crossover $\langle  n\rangle -T$ diagram highlighting different behaviors in the doped Holstein model for fixed $\lambda = 0.375$ and $\Omega/t=2.0$. Results are shown for 
    (a) the single-particle density of states at the Fermi energy $N(0)$, (b) the dc conductivity $\sigma_\mathrm{dc}(0)$ [see Eq.~\eqref{eq:sigma_dc}], (c) the uniform spin structure factor $S_\text{s}(\boldsymbol{q} = 0)$, and (d) charge structure factor $S_\text{c}(\boldsymbol{Q})$ at the CDW wave vector $\boldsymbol{Q}=(\pi,\pi)$. The solid and dashed lines indicate contours along which $N(0) = 0.005$ and $\sigma_\mathrm{dc} = 0.005$, respectively. } 
    \label{fig:crossover_20}
\end{figure}

We have carried out a similar analysis for other carrier concentrations, as summarized in Figs.~\ref{fig:crossover_05} and \ref{fig:crossover_20} for $\lambda = 0.375$ and $\Omega/t = 0.5$ and $2$, respectively. In both cases, we have overlaid contour lines to indicate different crossovers in the system's behavior (which should not be interpreted as phase transition lines). These lines mark constant values for different observables obtained from ridge-regression-smoothed splines of the data~\cite{Stickel2010Data}. In both figures, the solid lines indicate the  $N(0)=0.005$ contour while the dashed lines indicate the $\sigma_\mathrm{dc}=0.005$ contour. These small but nonzero values were chosen due to the noise level of the \gls*{DQMC} simulations and to overcome the difficulty of obtaining exactly zero in \gls*{AC}. Note that the dashed and solid contours in Fig.~\ref{fig:crossover_05} would approximately coincide, so we only show the solid contour in this case. 

The crossover diagrams indicate several regions of interest. The first is a gas of mobile carriers/polarons at small carrier concentrations, where the charge carriers are weakly dressed by the lattice and exhibit metallic conduction. This region is characterized by a high value of the single-particle density of states $N(0)$, the finite dc conductivity, a nonzero value of the uniform spin structure factor $S_\mathrm{s}(0)$, and a small charge structure factor $S_\mathrm{c}(\boldsymbol{Q})$. This region should have a transition to an $s$-wave \gls*{SC} at lower temperatures~\cite{Bradley2021Superconductivity, Nosarzewski2021superconductivity}. 
The second is a region of frozen bipolaron carriers, which generally appears at large carrier concentrations and low temperatures. It is characterized by a gap in the single-particle spectrum $N(0) = 0$, a suppressed uniform spin structure factor $S_\mathrm{s}(0)$ that is indicative of singlet formation, and zero dc conductivity. It also has a large $S_\mathrm{c}(\boldsymbol{Q})$, which grows toward half-filling and exhibits a large amount of noise, suggesting that the bipolarons have short-range $\boldsymbol{Q} = (\pi,\pi)$ order and possibly phase separation on short length scales. The filling fractions corresponding to the frozen bipolaron regime inferred coincide with the vertical regions in the $\langle n \rangle$ vs $\mu$ plots shown in Fig.~\ref{fig:Mu_vs_N}. Finally, for $\Omega/t = 2$, we also observe a small region hosting mobile bipolarons at large carrier concentrations and low temperatures. This region is characterized by a gap in the single-particle spectrum, $\sigma_\mathrm{dc}\ne0$, $S_\mathrm{s}(0) = 0$, and $S_\mathrm{c}(\boldsymbol{Q})\ne 0$. Notably, this region is absent in the $\Omega/t = 0.5$ case.  

\section{Discussion}\label{sec:discussion}

We have explored the spectral and transport properties of the \gls*{2D} square lattice Holstein model using nonperturbative \gls*{DQMC} together with different \gls*{AC} methods. By varying the model parameters, including the dimensionless \gls*{eph} coupling strength $\lambda$, phonon frequency $\Omega$, electronic filling $\langle n \rangle$, and temperature $T$, we systematically mapped out the model's electronic properties across large regions of parameter space, effectively elucidating a rich collection of ordering tendencies including \gls*{CDW} correlations and polaron formation. This study extended beyond prior systematic explorations of the Holstein model, both in the dilute \cite{Boncia1999holstein, Goodwin2006Greens, Ku2002dimensionality} and heavily doped~\cite{Bradley2021Superconductivity, Nosarzewski2021superconductivity} regimes, by providing results for its small polaron physics, including the transition from mobile (bi)polarons to immobile (self-trapped) bipolarons as the carrier concentration approaches half-filling. In general, we find that the self-trapping bipolaronic tendencies are enhanced in the adiabatic limit ($\Omega/E_\mathrm{F} < 1$). 

As mentioned in the introduction, \gls*{ARPES} measurements on correlated materials~\cite{Shen2004missing, Shen2007Doping, Mannella2005nodal, Mannella2007polaron} have observed broad Gaussian lineshapes, which have been associated with polaronic shake-off states similar to those derived in strong coupling expansions ~\cite{Alexandrov1992polaronic, Ranninger1993spectral}. The spectral functions observed here, however, do not develop such features; instead, we observe at most one or two clearly defined spectral features below the phonon energy rather than the expected ladder of phonon replicas. A recent \gls*{DMRG} study of the \gls*{1D} Hubbard-Holstein model at low electron densities also reported a similar lack of shake-off states~\cite{Kovac2025Signature}. We posit that a likely source for this discrepancy is the relative strength of the \gls*{eph} coupling. Having focused on $\lambda \le 0.5$ regime, we already find that bipolaron formation and self-trapping occur at these coupling strengths. In contrast, approximate theoretical treatments based on the Lang-Firsov transformation often begin from a quasi-localized picture with a heavily renormalized polaron band, implying a much larger effective coupling~\cite{Alexandrov1992polaronic, Ranninger1993spectral}. 

Discrepancies between the model and experimental observations could also arise from the simplicity of the Holstein Hamiltonian. For example, strong correlations present in the real material will generally suppress bipolaron formation and enhance the strength of the \gls*{eph} coupling~\cite{Huang2003electron, Yin2013correlation}. Phonon dispersion and other effects that should be present in real systems are known to strongly influence polaron properties in the dilute limit~\cite{Bonica2021dynamic}. Finally, nonlinear \gls*{eph} coupling~\cite{Dee2020relative, Adolphs2013going} and lattice anharmonicity~\cite{Paleari2021quantum} have also been shown to be relevant in the small polaron regime, where lattice distortions can become large enough to violate standard linear and harmonic approximations, and should also be considered when comparing to real materials. 

Finally, we have focused on the system's normal-state properties throughout this study. Prior \gls*{DQMC} studies of the Holstein model have already shown that polaron formation tends to suppress superconducting instabilities~\cite{Esterlis2018breakdown, Nosarzewski2021superconductivity}. We therefore expect any \gls*{SC} transitions to appear at lower temperatures than those accessed here and will generally be restricted to lower carrier concentrations, where (bi)polaron formation is not driving the system toward insulating behavior. This expectation is consistent with \gls*{DQMC} calculations that have only observed \gls*{SC} transitions at $\langle n \rangle \lessapprox 0.5-0.6$ and large phonon energies~\cite{Bradley2021Superconductivity, Nosarzewski2021spectral, Dee2020relative}. In the future, it would be interesting to carry out similar studies for other models, including the bond~\cite{Sengupta2003Peierls}, optical~\cite{Capone1997small}, and acoustic~\cite{Barisic1970tightbinding} \gls*{SSH} models. These types of models host light polarons and strong carrier binding~\cite{Sous2018light} whose fingerprints can be found in the single-particle spectral function at low doping levels~\cite{Banerjee2025spectral}, and would provide a more comprehensive view of how (bi)polaron formation can suppress superconductivity. These studies would also provide insights into recent proposals for high-$T_\mathrm{c}$ superconductivity in the bond \gls*{SSH} model~\cite{Zhang2023bipolaronic, TanjaroonLy2023comparative}.

\begin{acknowledgments}
We thank M. Berciu, P. M. Dee, and R. T. Scalettar for useful discussions and comments on this manuscript. This work was supported by the U.S.~Department of Energy, Office of Science, Office of Basic Energy Sciences, under Award Number DE-SC0022311. 
\end{acknowledgments}

\section*{Code Availability}\label{AppendixCode}
The \texttt{SmoQyDQMC.jl} and \texttt{SmoQyDEAC.jl} codes used in this study can be obtained from the SmoQy GitHub organization (\url{https://github.com/SmoQySuite/}). 

\section*{Data Availability}\label{AppendixData}
All data and scripts used in this paper have been deposited in an online repository~\cite{PaperRepo1}. Additional spectral data and visualization tools can be downloaded once all \gls*{AC} calculations have completed~\cite{PaperRepo2}.

\begin{appendix}
\section{Parameter Sets}\label{AppendixParameters}

\begin{table}[b]
\caption{\label{tab:parameters}Summary of the simulation parameters that were used to measure observables.}\hspace{0.25cm}
\begin{tabular}{ccc}
\hline
\multicolumn{1}{c}{\textbf{Parameter}}                       & \multicolumn{1}{c}{\textbf{Range}}                             & \multicolumn{1}{c}{\textbf{Step Size}}                  \\ \hline
$\langle n\rangle$                                           & 0.05-1                                                         & 0.05                                                    \\
$\Omega/t$                                                   & 0.5-2                                                          & 0.5                                                     \\
\begin{tabular}[c]{@{}c@{}}$\lambda$\\ $\lambda$ (additional)\end{tabular} & \begin{tabular}[c]{@{}c@{}}0.3-0.5\\ 0.125-0.75\end{tabular} & \begin{tabular}[c]{@{}c@{}}0.025\\ 0.125\end{tabular} \\
$\beta t$                                                    & 5-20                                                           & 1                                                       \\ \hline
\end{tabular}
\end{table}

Our \gls*{DQMC} simulations were carried out for 16,640 different parameter sets, as summarized in Tbl.~\ref{tab:parameters}. We also carried out an additional ``high-quality'' simulations that were used to compute the single-particle spectral functions and optical conductivity. (These parameter values are indicated as ``additional'' in Tbl.~\ref{tab:parameters}.) These data sets have been made publicly available (see Data Availability).

\section{Finite Size Effects}\label{AppendixFinite}
\begin{figure}[t]
    \centering
    \vspace{0.5cm}
    \includegraphics[width=0.8\linewidth]{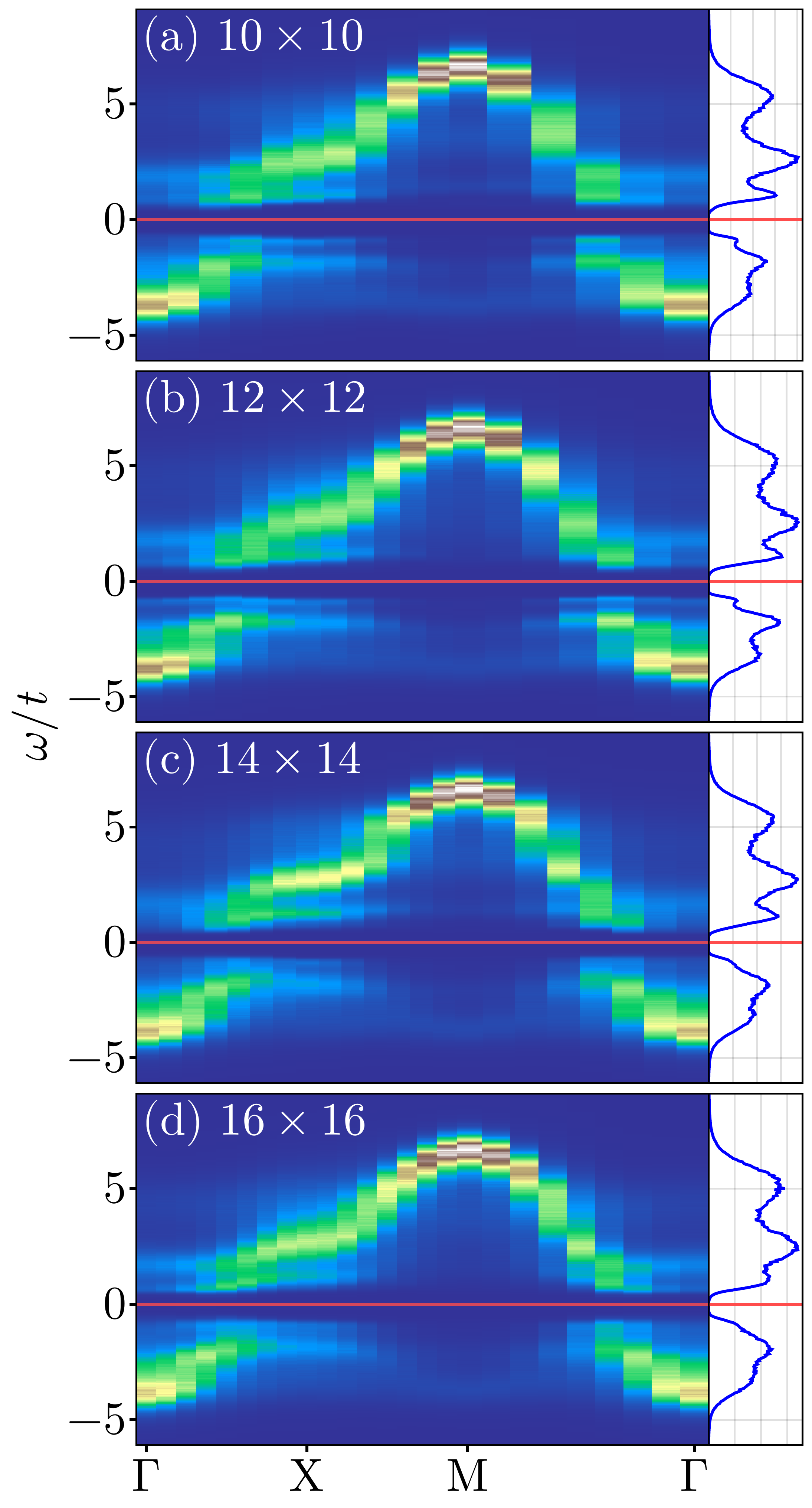}
    \caption{The single-particle spectral function along the high-symmetry cuts of the 
    first Brillouin zone. Results were obtained for $\langle n \rangle=0.6$, $\Omega/t=0.5$, $\lambda=0.375$, and $\beta t=20$ and on $N = L\times L$ clusters with (a) $L = 10$, (b) $12$, (c) $14$ and (d) $16$. The right column also shows the corresponding single-particle densities of states. }
    \label{fig:FiniteSize}
\end{figure}

Throughout the text, we have shown results obtained on $N = L\times L$ clusters with $L= 14$. While these sizes are fairly large by modern standards, it is important to assess potential finite-size effects, particularly at low temperatures where thermal broadening is relatively small. To this end, Fig.~\ref{fig:FiniteSize} plots the $A(\boldsymbol{k},\omega)$ along the high-symmetry cuts of the first Brillouin zone for different lattice sizes and a representative case of $\langle n \rangle=0.6$, $\Omega/t=0.5$, $\lambda=0.375$, and $\beta t=20$. The results across the different lattice sizes are consistent in that all four exhibit a large gap at the Fermi level and have additional kink-like band renormalizations indicative of \gls*{eph} coupling.  In all four cases, we also resolve sharp quasiparticle peaks near the $X$ point at $\omega\approx\pm t$ together with a classic peak-dip-hump structure. No clear indications of a ladder of phonon replica bands are visible near the $\Gamma$ point for any of the four systems.

\end{appendix}


\bibliography{references}

@article{White1989numerical,
  title = {Numerical study of the two-dimensional {H}ubbard model},
  author = {White, S. R. and Scalapino, D. J. and Sugar, R. L. and Loh, E. Y. and Gubernatis, J. E. and Scalettar, R. T.},
  journal = {Phys. Rev. B},
  volume = {40},
  issue = {1},
  pages = {506--516},
  numpages = {0},
  year = {1989},
  month = {Jul},
  publisher = {American Physical Society},
  doi = {10.1103/PhysRevB.40.506},
  url = {https://link.aps.org/doi/10.1103/PhysRevB.40.506}
}

@article{Khazraie2018oxygen,
  title = {Oxygen holes and hybridization in the bismuthates},
  author = {Khazraie, Arash and Foyevtsova, Kateryna and Elfimov, Ilya and Sawatzky, George A.},
  journal = {Phys. Rev. B},
  volume = {97},
  issue = {7},
  pages = {075103},
  numpages = {9},
  year = {2018},
  month = {Feb},
  publisher = {American Physical Society},
  doi = {10.1103/PhysRevB.97.075103},
  url = {https://link.aps.org/doi/10.1103/PhysRevB.97.075103}
}

@Article{SmoqyDQMC2,
	title={{Codebase release r2.0 for SmoQyDQMC.jl}},
	author={Benjamin Cohen-Stead and Shruti Agarwal and Sohan Malkaruge Costa and James Neuhaus and Andy Tanjaroon Ly and Yutan Zhang and Richard Scalettar and Kipton Barros and Steven Johnston},
	journal={SciPost Phys. Codebases},
	pages={29-r2.0},
	year={2026},
	publisher={SciPost},
	doi={10.21468/SciPostPhysCodeb.29-r2.0},
	url={https://scipost.org/10.21468/SciPostPhysCodeb.29-r2.0},
}

@Article{SmoqyDQMC1,
	title={{SmoQyDQMC.jl: A flexible implementation of determinant quantum {M}onte {C}arlo for {H}ubbard and electron-phonon interactions (version 2.0 release)}},
	author={Benjamin Cohen-Stead and Shruti Agarwal and Sohan Malkaruge Costa and James Neuhaus and Andy Tanjaroon Ly and Yutan Zhang and Richard Scalettar and Kipton Barros and Steven Johnston},
	journal={SciPost Phys. Codebases},
	pages={29-v2.0},
	year={2026},
	publisher={SciPost},
	doi={10.21468/SciPostPhysCodeb.29-v2.0},
	url={https://scipost.org/10.21468/SciPostPhysCodeb.29-v2.0},
}

@article{Miles2022dynamical,
  title = {Dynamical tuning of the chemical potential to achieve a target particle number in grand canonical {M}onte {C}arlo simulations},
  author = {Miles, Cole and Cohen-Stead, Benjamin and Bradley, Owen and Johnston, Steven and Scalettar, Richard and Barros, Kipton},
  journal = {Phys. Rev. E},
  volume = {105},
  issue = {4},
  pages = {045311},
  numpages = {9},
  year = {2022},
  month = {Apr},
  publisher = {American Physical Society},
  doi = {10.1103/PhysRevE.105.045311},
  url = {https://link.aps.org/doi/10.1103/PhysRevE.105.045311}
}

@Article{Neuhaus2024SmoQyDEAC1,
	title={{SmoQyDEAC.jl: A differential evolution package for the analytic continuation of imaginary time correlation functions}},
	author={James Neuhaus and Nathan S. Nichols and Debshikha Banerjee and Benjamin Cohen-Stead and Thomas A. Maier and Adrian Del Maestro and Steven Johnston},
	journal={SciPost Phys. Codebases},
	pages={39},
	year={2024},
	publisher={SciPost},
	doi={10.21468/SciPostPhysCodeb.39},
	url={https://scipost.org/10.21468/SciPostPhysCodeb.39},
}

@Article{Neuhaus2024SmoQyDEAC2,
	title={{Codebase release r1.1 for SmoQyDEAC.jl}},
	author={James Neuhaus and Nathan S. Nichols and Debshikha Banerjee and Benjamin Cohen-Stead and Thomas A. Maier and Adrian Del Maestro and Steven Johnston},
	journal={SciPost Phys. Codebases},
	pages={39-r1.1},
	year={2024},
	publisher={SciPost},
	doi={10.21468/SciPostPhysCodeb.39-r1.1},
	url={https://scipost.org/10.21468/SciPostPhysCodeb.39-r1.1},
}

@article{Nichols2022Parameter,
  title = {Parameter-free differential evolution algorithm for the analytic continuation of imaginary time correlation functions},
  author = {Nichols, Nathan S. and Sokol, Paul and Del Maestro, Adrian},
  journal = {Phys. Rev. E},
  volume = {106},
  issue = {2},
  pages = {025312},
  numpages = {10},
  year = {2022},
  month = {Aug},
  publisher = {American Physical Society},
  doi = {10.1103/PhysRevE.106.025312},
  url = {https://link.aps.org/doi/10.1103/PhysRevE.106.025312}
}

@article{Scalettar1999Quantum,
  title = {Quantum {M}onte {C}arlo study of the disordered attractive {H}ubbard model},
  author = {Scalettar, R. T. and Trivedi, N. and Huscroft, C.},
  journal = {Phys. Rev. B},
  volume = {59},
  issue = {6},
  pages = {4364--4375},
  numpages = {0},
  year = {1999},
  month = {Feb},
  publisher = {American Physical Society},
  doi = {10.1103/PhysRevB.59.4364},
  url = {https://link.aps.org/doi/10.1103/PhysRevB.59.4364}
}

@article{Schattner2016Ising,
  title = {Ising Nematic Quantum Critical Point in a Metal: A {M}onte {C}arlo Study},
  author = {Schattner, Yoni and Lederer, Samuel and Kivelson, Steven A. and Berg, Erez},
  journal = {Phys. Rev. X},
  volume = {6},
  issue = {3},
  pages = {031028},
  numpages = {18},
  year = {2016},
  month = {Aug},
  publisher = {American Physical Society},
  doi = {10.1103/PhysRevX.6.031028},
  url = {https://link.aps.org/doi/10.1103/PhysRevX.6.031028}
}

@article{Issa2025learning,
  title = {Learning by confusion: The phase diagram of the {H}olstein model},
  author = {Issa, George and Bradley, Owen and Khatami, Ehsan and Scalettar, Richard},
  journal = {Phys. Rev. B},
  volume = {111},
  issue = {15},
  pages = {155140},
  numpages = {10},
  year = {2025},
  month = {Apr},
  publisher = {American Physical Society},
  doi = {10.1103/PhysRevB.111.155140},
  url = {https://link.aps.org/doi/10.1103/PhysRevB.111.155140}
}

@article{Marsiglio1990pairing,
  title = {Pairing and charge-density-wave correlations in the {H}olstein model at half-filling},
  author = {Marsiglio, F.},
  journal = {Phys. Rev. B},
  volume = {42},
  issue = {4},
  pages = {2416--2424},
  numpages = {0},
  year = {1990},
  month = {Aug},
  publisher = {American Physical Society},
  doi = {10.1103/PhysRevB.42.2416},
  url = {https://link.aps.org/doi/10.1103/PhysRevB.42.2416}
}

@article{Franchini2021polarons,
	author = {Franchini, Cesare and Reticcioli, Michele and Setvin, Martin and Diebold, Ulrike},
	date = {2021/07/01},
	doi = {10.1038/s41578-021-00289-w},
	id = {Franchini2021},
	isbn = {2058-8437},
	journal = {Nature Reviews Materials},
	number = {7},
	pages = {560--586},
	title = {Polarons in materials},
	url = {https://doi.org/10.1038/s41578-021-00289-w},
	volume = {6},
	year = {2021}}

@article{Allodi2001Ultraslow, 
    title = {Ultraslow Polaron Dynamics in Low-Doped Manganites from {$^{139}$La NMR-NQR} and Muon Spin Rotation},
  author = {Allodi, G. and Cestelli Guidi, M. and De Renzi, R. and Caneiro, A. and Pinsard, L.},
  journal = {Phys. Rev. Lett.},
  volume = {87},
  issue = {12},
  pages = {127206},
  numpages = {4},
  year = {2001},
  month = {Aug},
  publisher = {American Physical Society},
  doi = {10.1103/PhysRevLett.87.127206},
  url = {https://link.aps.org/doi/10.1103/PhysRevLett.87.127206}
}

@article{Swartz2018polaronic, 
author = {Adrian G. Swartz  and Hisashi Inoue  and Tyler A. Merz  and Yasuyuki Hikita  and Srinivas Raghu  and Thomas P. Devereaux  and Steven Johnston  and Harold Y. Hwang },
title = {Polaronic behavior in a weak-coupling superconductor},
journal = {Proceedings of the National Academy of Sciences},
volume = {115},
number = {7},
pages = {1475-1480},
year = {2018},
doi = {10.1073/pnas.1713916115},
URL = {https://www.pnas.org/doi/abs/10.1073/pnas.1713916115}}

@article{Frederikse1964electronic,
  title = {Electronic Transport in Strontium Titanate},
  author = {Frederikse, H. P. R. and Thurber, W. R. and Hosler, W. R.},
  journal = {Phys. Rev.},
  volume = {134},
  issue = {2A},
  pages = {A442--A445},
  numpages = {0},
  year = {1964},
  month = {Apr},
  publisher = {American Physical Society},
  doi = {10.1103/PhysRev.134.A442},
  url = {https://link.aps.org/doi/10.1103/PhysRev.134.A442}
}

@article{Yang2013intrinsic,
  title = {Intrinsic small polarons in rutile {TiO$_{2}$}},
  author = {Yang, Shan and Brant, A. T. and Giles, N. C. and Halliburton, L. E.},
  journal = {Phys. Rev. B},
  volume = {87},
  issue = {12},
  pages = {125201},
  numpages = {6},
  year = {2013},
  month = {Mar},
  publisher = {American Physical Society},
  doi = {10.1103/PhysRevB.87.125201},
  url = {https://link.aps.org/doi/10.1103/PhysRevB.87.125201}
}

@article{Bredas2985polarons,
	author = {Bredas, Jean Luc and Street, G. Bryan},
	date = {1985/10/01},
	doi = {10.1021/ar00118a005},
	isbn = {0001-4842},
	journal = {Accounts of Chemical Research},
	journal1 = {Accounts of Chemical Research},
	journal2 = {Acc. Chem. Res.},
	month = {10},
	number = {10},
	pages = {309--315},
	publisher = {American Chemical Society},
	title = {Polarons, bipolarons, and solitons in conducting polymers},
	type = {doi: 10.1021/ar00118a005},
	url = {https://doi.org/10.1021/ar00118a005},
	volume = {18},
	year = {1985},
	year1 = {1985}}

@book{Alexandrov2010,
	address = {Berlin, Heidelberg},
	author = {Alexandrov, Alexandre S. and Devreese, Jozef T.},
	title = {Advances in Polaron Physics},
	doi = {10.1007/978-3-642-01896-1_1},
	isbn = {978-3-642-01896-1},
	pages = {1--9},
	publisher = {Springer Berlin Heidelberg},
	url = {https://doi.org/10.1007/978-3-642-01896-1_1},
	year = {2010}}

@article{Marsiglio1995signatures,
  title = {Signatures of the electron-phonon interaction in the far-infrared},
  author = {Marsiglio, F. and Carbotte, J. P.},
  journal = {Phys. Rev. B},
  volume = {52},
  issue = {22},
  pages = {16192--16198},
  numpages = {0},
  year = {1995},
  month = {Dec},
  publisher = {American Physical Society},
  doi = {10.1103/PhysRevB.52.16192},
  url = {https://link.aps.org/doi/10.1103/PhysRevB.52.16192}
}

@article{Engelsberg1963coupled,
  title = {Coupled Electron-Phonon System},
  author = {Engelsberg, S. and Schrieffer, J. R.},
  journal = {Phys. Rev.},
  volume = {131},
  issue = {3},
  pages = {993--1008},
  numpages = {0},
  year = {1963},
  month = {Aug},
  publisher = {American Physical Society},
  doi = {10.1103/PhysRev.131.993},
  url = {https://link.aps.org/doi/10.1103/PhysRev.131.993}
}

@article{Shao2023progress,
	author = {Hui Shao and Anders W. Sandvik},
	doi = {https://doi.org/10.1016/j.physrep.2022.11.002},
	issn = {0370-1573},
	journal = {Physics Reports},
	pages = {1-88},
	title = {Progress on stochastic analytic continuation of quantum {M}onte {C}arlo data},
	url = {https://www.sciencedirect.com/science/article/pii/S0370157322003921},
	volume = {1003},
	year = {2023}}

@article{Dee2019temperature,
  title = {Temperature-filling phase diagram of the two-dimensional {H}olstein model in the thermodynamic limit by self-consistent {M}igdal approximation},
  author = {Dee, P. M. and Nakatsukasa, K. and Wang, Y. and Johnston, S.},
  journal = {Phys. Rev. B},
  volume = {99},
  issue = {2},
  pages = {024514},
  numpages = {18},
  year = {2019},
  month = {Jan},
  publisher = {American Physical Society},
  doi = {10.1103/PhysRevB.99.024514},
  url = {https://link.aps.org/doi/10.1103/PhysRevB.99.024514}
}

@article{Esterlis2018breakdown,
  title = {Breakdown of the {M}igdal-{E}liashberg theory: A determinant quantum {M}onte {C}arlo study},
  author = {Esterlis, I. and Nosarzewski, B. and Huang, E. W. and Moritz, B. and Devereaux, T. P. and Scalapino, D. J. and Kivelson, S. A.},
  journal = {Phys. Rev. B},
  volume = {97},
  issue = {14},
  pages = {140501},
  numpages = {5},
  year = {2018},
  month = {Apr},
  publisher = {American Physical Society},
  doi = {10.1103/PhysRevB.97.140501},
  url = {https://link.aps.org/doi/10.1103/PhysRevB.97.140501}
}

@article{Nosarzewski2021superconductivity,
  title = {Superconductivity, charge density waves, and bipolarons in the {H}olstein model},
  author = {Nosarzewski, B. and Huang, E. W. and Dee, Philip M. and Esterlis, I. and Moritz, B. and Kivelson, S. A. and Johnston, S. and Devereaux, T. P.},
  journal = {Phys. Rev. B},
  volume = {103},
  issue = {23},
  pages = {235156},
  numpages = {9},
  year = {2021},
  month = {Jun},
  publisher = {American Physical Society},
  doi = {10.1103/PhysRevB.103.235156},
  url = {https://link.aps.org/doi/10.1103/PhysRevB.103.235156}
}

@article{Stickel2010Data,
title = {Data smoothing and numerical differentiation by a regularization method},
journal = {Computers \& Chemical Engineering},
volume = {34},
number = {4},
pages = {467-475},
year = {2010},
issn = {0098-1354},
doi = {https://doi.org/10.1016/j.compchemeng.2009.10.007},
url = {https://www.sciencedirect.com/science/article/pii/S0098135409002567},
author = {Jonathan J. Stickel}
}

@article{Bonica2021dynamic,
  title = {Dynamic properties of a polaron coupled to dispersive optical phonons},
  author = {Bon\ifmmode \check{c}\else \v{c}\fi{}a, J. and Trugman, S. A.},
  journal = {Phys. Rev. B},
  volume = {103},
  issue = {5},
  pages = {054304},
  numpages = {9},
  year = {2021},
  month = {Feb},
  publisher = {American Physical Society},
  doi = {10.1103/PhysRevB.103.054304},
  url = {https://link.aps.org/doi/10.1103/PhysRevB.103.054304}
}

@article{Huang2003electron,
  title = {Electron-phonon vertex in the two-dimensional one-band {H}ubbard model},
  author = {Huang, Z. B. and Hanke, W. and Arrigoni, E. and Scalapino, D. J.},
  journal = {Phys. Rev. B},
  volume = {68},
  issue = {22},
  pages = {220507(R)},
  numpages = {4},
  year = {2003},
  month = {Dec},
  publisher = {American Physical Society},
  doi = {10.1103/PhysRevB.68.220507},
  url = {https://link.aps.org/doi/10.1103/PhysRevB.68.220507}
}

@article{Yin2013correlation,
  title = {Correlation-Enhanced Electron-Phonon Coupling: Applications of {$GW$} and Screened Hybrid Functional to Bismuthates, Chloronitrides, and Other High-${T}_{c}$ Superconductors},
  author = {Yin, Z. P. and Kutepov, A. and Kotliar, G.},
  journal = {Phys. Rev. X},
  volume = {3},
  issue = {2},
  pages = {021011},
  numpages = {20},
  year = {2013},
  month = {May},
  publisher = {American Physical Society},
  doi = {10.1103/PhysRevX.3.021011},
  url = {https://link.aps.org/doi/10.1103/PhysRevX.3.021011}
}

@article{Jarrell1996Bayesian,
title = {Bayesian inference and the analytic continuation of imaginary-time quantum {M}onte {C}arlo data},
journal = {Physics Reports},
volume = {269},
number = {3},
pages = {133-195},
year = {1996},
issn = {0370-1573},
doi = {https://doi.org/10.1016/0370-1573(95)00074-7},
url = {https://www.sciencedirect.com/science/article/pii/0370157395000747},
author = {Mark Jarrell and J.E. Gubernatis}
}

@article{Cheng2023Machine,
  title = {Machine learning for phase ordering dynamics of charge density waves},
  author = {Cheng, Chen and Zhang, Sheng and Chern, Gia-Wei},
  journal = {Phys. Rev. B},
  volume = {108},
  issue = {1},
  pages = {014301},
  numpages = {15},
  year = {2023},
  month = {Jul},
  publisher = {American Physical Society},
  doi = {10.1103/PhysRevB.108.014301},
  url = {https://link.aps.org/doi/10.1103/PhysRevB.108.014301}
}

@article{Huang2019Strange,
author = {Edwin W. Huang  and Ryan Sheppard  and Brian Moritz  and Thomas P. Devereaux },
title = {Strange metallicity in the doped {H}ubbard model},
journal = {Science},
volume = {366},
number = {6468},
pages = {987-990},
year = {2019},
doi = {10.1126/science.aau7063},
URL = {https://www.science.org/doi/abs/10.1126/science.aau7063}}

@article{Bergeron2016Algorithms,
  title = {Algorithms for optimized maximum entropy and diagnostic tools for analytic continuation},
  author = {Bergeron, Dominic and Tremblay, A.-M. S.},
  journal = {Phys. Rev. E},
  volume = {94},
  issue = {2},
  pages = {023303},
  numpages = {25},
  year = {2016},
  month = {Aug},
  publisher = {American Physical Society},
  doi = {10.1103/PhysRevE.94.023303},
  url = {https://link.aps.org/doi/10.1103/PhysRevE.94.023303}
}

@article{Scalettar1989Competition,
  title = {Competition of pairing and {P}eierls--charge-density-wave correlations in a two-dimensional electron-phonon model},
  author = {Scalettar, R. T. and Bickers, N. E. and Scalapino, D. J.},
  journal = {Phys. Rev. B},
  volume = {40},
  issue = {1},
  pages = {197--200},
  numpages = {0},
  year = {1989},
  month = {Jul},
  publisher = {American Physical Society},
  doi = {10.1103/PhysRevB.40.197},
  url = {https://link.aps.org/doi/10.1103/PhysRevB.40.197}
}

@article{Holstein1959Studies,
title = {Studies of polaron motion: Part {I}. The molecular-crystal model},
journal = {Annals of Physics},
volume = {8},
number = {3},
pages = {325-342},
year = {1959},
issn = {0003-4916},
doi = {https://doi.org/10.1016/0003-4916(59)90002-8},
url = {https://www.sciencedirect.com/science/article/pii/0003491659900028},
author = {T. Holstein}
}

@article{Menushenkov2024Direct,
  title = {Direct evidence of real-space pairing in {$\mathrm{Ba}{\mathrm{BiO}}_{3}$}},
  author = {Menushenkov, A. P. and Ivanov, A. and Neverov, V. and Lukyanov, A. and Krasavin, A. and Yastrebtsev, A. A. and Kovalev, I. A. and Zhumagulov, Y. and Kuznetsov, A. V. and Popov, V. and Tselikov, G. and Shchetinin, I. and Krymskaya, O. and Yaroslavtsev, A. and Carley, R. and Mercadier, L. and Yin, Z. and Parchenko, S. and Hoang, L. P. and Ghodrati, N. and Kim, Y. Y. and Schlappa, J. and Izquierdo, M. and Molodtsov, S. and Scherz, A.},
  journal = {Phys. Rev. Res.},
  volume = {6},
  issue = {2},
  pages = {023307},
  numpages = {14},
  year = {2024},
  month = {Jun},
  publisher = {American Physical Society},
  doi = {10.1103/PhysRevResearch.6.023307},
  url = {https://link.aps.org/doi/10.1103/PhysRevResearch.6.023307}
}

@article{Schirmer1980Conduction,
doi = {10.1088/0022-3719/13/36/005},
url = {https://dx.doi.org/10.1088/0022-3719/13/36/005},
year = {1980},
month = {dec},
publisher = {},
volume = {13},
number = {36},
pages = {L1067},
author = {O F Schirmer and E Salje},
title = {Conduction bipolarons in low-temperature crystalline {WO$_{3-x}$}},
journal = {Journal of Physics C: Solid State Physics}
}

@Inbook{Shen2007Doping,
author="Shen, K. M.
and Shen, Z.-X.",
editor="H{\"u}fner, Stefan",
title="Doping Evolution of the Cuprate Superconductors from High-Resolution {ARPES}",
bookTitle="Very High Resolution Photoelectron Spectroscopy",
year="2007",
publisher="Springer Berlin Heidelberg",
address="Berlin, Heidelberg",
pages="243--270",
isbn="978-3-540-68133-5",
doi="10.1007/3-540-68133-7_9",
url="https://doi.org/10.1007/3-540-68133-7_9"
}

@article{Sengupta2003Peierls,
  title = {Peierls transition in the presence of finite-frequency phonons in the one-dimensional extended {P}eierls-{H}ubbard model at half-filling},
  author = {Sengupta, Pinaki and Sandvik, Anders W. and Campbell, David K.},
  journal = {Phys. Rev. B},
  volume = {67},
  issue = {24},
  pages = {245103},
  numpages = {6},
  year = {2003},
  month = {Jun},
  publisher = {American Physical Society},
  doi = {10.1103/PhysRevB.67.245103},
  url = {https://link.aps.org/doi/10.1103/PhysRevB.67.245103}
}

@article{Capone1997small,
  title = {Small-polaron formation and optical absorption in {S}u-{S}chrieffer-{H}eeger and {H}olstein models},
  author = {Capone, M. and Stephan, W. and Grilli, M.},
  journal = {Phys. Rev. B},
  volume = {56},
  issue = {8},
  pages = {4484--4493},
  numpages = {0},
  year = {1997},
  month = {Aug},
  publisher = {American Physical Society},
  doi = {10.1103/PhysRevB.56.4484},
  url = {https://link.aps.org/doi/10.1103/PhysRevB.56.4484}
}

@article{Banerjee2025spectral,
  title = {Spectral signatures of residual electron pairing in the extended {H}ubbard-{S}u-{S}chrieffer-{H}eeger model},
  author = {Banerjee, Debshikha and Nocera, Alberto and Sawatzky, George A. and Berciu, Mona and Johnston, Steven},
  journal = {Phys. Rev. B},
  volume = {111},
  issue = {24},
  pages = {L241107},
  numpages = {6},
  year = {2025},
  month = {Jun},
  publisher = {American Physical Society},
  doi = {10.1103/PhysRevB.111.L241107},
  url = {https://link.aps.org/doi/10.1103/PhysRevB.111.L241107}
}

@article{Sous2018light,
  title = {Light Bipolarons Stabilized by {P}eierls Electron-Phonon Coupling},
  author = {Sous, John and Chakraborty, Monodeep and Krems, Roman V. and Berciu, Mona},
  journal = {Phys. Rev. Lett.},
  volume = {121},
  issue = {24},
  pages = {247001},
  numpages = {6},
  year = {2018},
  month = {Dec},
  publisher = {American Physical Society},
  doi = {10.1103/PhysRevLett.121.247001},
  url = {https://link.aps.org/doi/10.1103/PhysRevLett.121.247001}
}

@article{Zhang2023bipolaronic,
  title = {Bipolaronic High-Temperature Superconductivity},
  author = {Zhang, C. and Sous, J. and Reichman, D. R. and Berciu, M. and Millis, A. J. and Prokof'ev, N. V. and Svistunov, B. V.},
  journal = {Phys. Rev. X},
  volume = {13},
  issue = {1},
  pages = {011010},
  numpages = {19},
  year = {2023},
  month = {Jan},
  publisher = {American Physical Society},
  doi = {10.1103/PhysRevX.13.011010},
  url = {https://link.aps.org/doi/10.1103/PhysRevX.13.011010}
}

@article{Calvani1996Infrared,
  title = {Infrared response of ordered polarons in layered perovskites},
  author = {Calvani, P. and Paolone, A. and Dore, P. and Lupi, S. and Maselli, P. and Medaglia, P. G. and Cheong, S -W.},
  journal = {Phys. Rev. B},
  volume = {54},
  issue = {14},
  pages = {R9592--R9595},
  numpages = {0},
  year = {1996},
  month = {Oct},
  publisher = {American Physical Society},
  doi = {10.1103/PhysRevB.54.R9592},
  url = {https://link.aps.org/doi/10.1103/PhysRevB.54.R9592}
}

@article{Nosarzewski2021spectral,
  title = {Spectral properties and enhanced superconductivity in renormalized {M}igdal-{E}liashberg theory},
  author = {Nosarzewski, Benjamin and Sch\"uler, Michael and Devereaux, Thomas P.},
  journal = {Phys. Rev. B},
  volume = {103},
  issue = {2},
  pages = {024520},
  numpages = {10},
  year = {2021},
  month = {Jan},
  publisher = {American Physical Society},
  doi = {10.1103/PhysRevB.103.024520},
  url = {https://link.aps.org/doi/10.1103/PhysRevB.103.024520}
}

@article{Mitric2022spectral,
  title = {Spectral Functions of the {H}olstein Polaron: Exact and Approximate Solutions},
  author = {Mitri\ifmmode \acute{c}\else \'{c}\fi{}, Petar and Jankovi\ifmmode \acute{c}\else \'{c}\fi{}, Veljko and Vukmirovi\ifmmode \acute{c}\else \'{c}\fi{}, Nenad and Tanaskovi\ifmmode \acute{c}\else \'{c}\fi{}, Darko},
  journal = {Phys. Rev. Lett.},
  volume = {129},
  issue = {9},
  pages = {096401},
  numpages = {6},
  year = {2022},
  month = {Aug},
  publisher = {American Physical Society},
  doi = {10.1103/PhysRevLett.129.096401},
  url = {https://link.aps.org/doi/10.1103/PhysRevLett.129.096401}
}

@article{Karakuzu2017superconductivity,
  title = {Superconductivity, charge-density waves, antiferromagnetism, and phase separation in the {H}ubbard-{H}olstein model},
  author = {Karakuzu, Seher and Tocchio, Luca F. and Sorella, Sandro and Becca, Federico},
  journal = {Phys. Rev. B},
  volume = {96},
  issue = {20},
  pages = {205145},
  numpages = {10},
  year = {2017},
  month = {Nov},
  publisher = {American Physical Society},
  doi = {10.1103/PhysRevB.96.205145},
  url = {https://link.aps.org/doi/10.1103/PhysRevB.96.205145}
}

@article{Nocera2014interplay,
  title = {Interplay of charge, spin, and lattice degrees of freedom in the spectral properties of the one-dimensional {H}ubbard-{H}olstein model},
  author = {Nocera, A. and Soltanieh-ha, M. and Perroni, C. A. and Cataudella, V. and Feiguin, A. E.},
  journal = {Phys. Rev. B},
  volume = {90},
  issue = {19},
  pages = {195134},
  numpages = {9},
  year = {2014},
  month = {Nov},
  publisher = {American Physical Society},
  doi = {10.1103/PhysRevB.90.195134},
  url = {https://link.aps.org/doi/10.1103/PhysRevB.90.195134}
}

@article{CohenStead2020langevin,
  title = {Langevin simulations of the half-filled cubic {H}olstein model},
  author = {Cohen-Stead, B. and Barros, Kipton and Meng, ZY and Chen, Chuang and Scalettar, R. T. and Batrouni, G. G.},
  journal = {Phys. Rev. B},
  volume = {102},
  issue = {16},
  pages = {161108(R)},
  numpages = {5},
  year = {2020},
  month = {Oct},
  publisher = {American Physical Society},
  doi = {10.1103/PhysRevB.102.161108},
  url = {https://link.aps.org/doi/10.1103/PhysRevB.102.161108}
}

@article{Naamneh2025persistence,
  title = {Persistence of small polarons into the superconducting doping range of {${\mathrm{Ba}}_{1\ensuremath{-}x}{\mathrm{K}}_{x}{\mathrm{BiO}}_{3}$}},
  author = {Naamneh, Muntaser and O'Quinn, Eric C. and Paris, Eugenio and McNally, Daniel and Tseng, Yi and Pude\l{}ko, Wojciech R. and Gawryluk, Dariusz J. and Shamblin, Jacob and Cohen-Stead, Benjamin and Shi, Ming and Radovic, Milan and Lang, Maik K. and Schmitt, Thorsten and Johnston, Steven and Plumb, Nicholas C.},
  journal = {Phys. Rev. Res.},
  volume = {7},
  issue = {4},
  pages = {043082},
  numpages = {13},
  year = {2025},
  month = {Oct},
  publisher = {American Physical Society},
  doi = {10.1103/s3p1-cy1s},
  url = {https://link.aps.org/doi/10.1103/s3p1-cy1s}
}

@article{Mannella2005nodal,
	author = {Mannella, N. and Yang, W. L. and Zhou, X. J. and Zheng, H. and Mitchell, J. F. and Zaanen, J. and Devereaux, T. P. and Nagaosa, N. and Hussain, Z. and Shen, Z. -X.},
	date = {2005/11/01},
	doi = {10.1038/nature04273},
	id = {Mannella2005},
	isbn = {1476-4687},
	journal = {Nature},
	number = {7067},
	pages = {474--478},
	title = {Nodal quasiparticle in pseudogapped colossal magnetoresistive manganites},
	url = {https://doi.org/10.1038/nature04273},
	volume = {438},
	year = {2005}}

@article{Mannella2007polaron,
  title = {Polaron coherence condensation as the mechanism for colossal magnetoresistance in layered manganites},
  author = {Mannella, N. and Yang, W. L. and Tanaka, K. and Zhou, X. J. and Zheng, H. and Mitchell, J. F. and Zaanen, J. and Devereaux, T. P. and Nagaosa, N. and Hussain, Z. and Shen, Z.-X.},
  journal = {Phys. Rev. B},
  volume = {76},
  issue = {23},
  pages = {233102},
  numpages = {4},
  year = {2007},
  month = {Dec},
  publisher = {American Physical Society},
  doi = {10.1103/PhysRevB.76.233102},
  url = {https://link.aps.org/doi/10.1103/PhysRevB.76.233102}
}

@article{Jooss2007polaron, 
author = {Ch. Jooss  and L. Wu  and T. Beetz  and R. F. Klie  and M. Beleggia  and M. A. Schofield  and S. Schramm  and J. Hoffmann  and Y. Zhu },
title = {Polaron melting and ordering as key mechanisms for colossal resistance effects in manganites},
journal = {Proceedings of the National Academy of Sciences},
volume = {104},
number = {34},
pages = {13597-13602},
year = {2007},
doi = {10.1073/pnas.0702748104},
URL = {https://www.pnas.org/doi/abs/10.1073/pnas.0702748104}}

@article{Ku2002dimensionality,
  title = {Dimensionality effects on the {H}olstein polaron},
  author = {Ku, Li-Chung and Trugman, S. A. and Bon\ifmmode \check{c}\else \v{c}\fi{}a, J.},
  journal = {Phys. Rev. B},
  volume = {65},
  issue = {17},
  pages = {174306},
  numpages = {10},
  year = {2002},
  month = {Apr},
  publisher = {American Physical Society},
  doi = {10.1103/PhysRevB.65.174306},
  url = {https://link.aps.org/doi/10.1103/PhysRevB.65.174306}
}

@article{Mishchenko2014diagrammatic,
  title = {Diagrammatic {M}onte {C}arlo Method for Many-Polaron Problems},
  author = {Mishchenko, Andrey S. and Nagaosa, Naoto and Prokof'ev, Nikolay},
  journal = {Phys. Rev. Lett.},
  volume = {113},
  issue = {16},
  pages = {166402},
  numpages = {5},
  year = {2014},
  month = {Oct},
  publisher = {American Physical Society},
  doi = {10.1103/PhysRevLett.113.166402},
  url = {https://link.aps.org/doi/10.1103/PhysRevLett.113.166402}
}

@article{Goodwin2006Greens,
  title = {Green's function of the {H}olstein polaron},
  author = {Goodvin, Glen L. and Berciu, Mona and Sawatzky, George A.},
  journal = {Phys. Rev. B},
  volume = {74},
  issue = {24},
  pages = {245104},
  numpages = {22},
  year = {2006},
  month = {Dec},
  publisher = {American Physical Society},
  doi = {10.1103/PhysRevB.74.245104},
  url = {https://link.aps.org/doi/10.1103/PhysRevB.74.245104}
}

@article{Boncia1999holstein,
  title = {{H}olstein polaron},
  author = {Bon\ifmmode \check{c}\else \v{c}\fi{}a, J. and Trugman, S. A. and Batisti\ifmmode \acute{c}\else \'{c}\fi{}, I.},
  journal = {Phys. Rev. B},
  volume = {60},
  issue = {3},
  pages = {1633--1642},
  numpages = {0},
  year = {1999},
  month = {Jul},
  publisher = {American Physical Society},
  doi = {10.1103/PhysRevB.60.1633},
  url = {https://link.aps.org/doi/10.1103/PhysRevB.60.1633}
}

@article{Shen2004missing,
  title = {Missing Quasiparticles and the Chemical Potential Puzzle in the Doping Evolution of the Cuprate Superconductors},
  author = {Shen, K. M. and Ronning, F. and Lu, D. H. and Lee, W. S. and Ingle, N. J. C. and Meevasana, W. and Baumberger, F. and Damascelli, A. and Armitage, N. P. and Miller, L. L. and Kohsaka, Y. and Azuma, M. and Takano, M. and Takagi, H. and Shen, Z.-X.},
  journal = {Phys. Rev. Lett.},
  volume = {93},
  issue = {26},
  pages = {267002},
  numpages = {4},
  year = {2004},
  month = {Dec},
  publisher = {American Physical Society},
  doi = {10.1103/PhysRevLett.93.267002},
  url = {https://link.aps.org/doi/10.1103/PhysRevLett.93.267002}
}

@article{Hohenadler2005photoemission,
  title = {Photoemission spectra of many-polaron systems},
  author = {Hohenadler, M. and Neuber, D. and von der Linden, W. and Wellein, G. and Loos, J. and Fehske, H.},
  journal = {Phys. Rev. B},
  volume = {71},
  issue = {24},
  pages = {245111},
  numpages = {15},
  year = {2005},
  month = {Jun},
  publisher = {American Physical Society},
  doi = {10.1103/PhysRevB.71.245111},
  url = {https://link.aps.org/doi/10.1103/PhysRevB.71.245111}
}

@article{Dee2020relative,
	author = {Dee, Philip M. and Coulter, Jennifer and Kleiner, Kevin G. and Johnston, Steven},
	date = {2020/08/21},
	doi = {10.1038/s42005-020-00413-2},
	id = {Dee2020},
	isbn = {2399-3650},
	journal = {Communications Physics},
	number = {1},
	pages = {145},
	title = {Relative importance of nonlinear electron-phonon coupling and vertex corrections in the {H}olstein model},
	url = {https://doi.org/10.1038/s42005-020-00413-2},
	volume = {3},
	year = {2020}}

@article{Mishchenko2008charge,
  title = {Charge Dynamics of Doped Holes in High {${T}_{c}$} Cuprate Superconductors: A Clue from Optical Conductivity},
  author = {Mishchenko, A. S. and Nagaosa, N. and Shen, Z.-X. and De Filippis, G. and Cataudella, V. and Devereaux, T. P. and Bernhard, C. and Kim, K. W. and Zaanen, J.},
  journal = {Phys. Rev. Lett.},
  volume = {100},
  issue = {16},
  pages = {166401},
  numpages = {4},
  year = {2008},
  month = {Apr},
  publisher = {American Physical Society},
  doi = {10.1103/PhysRevLett.100.166401},
  url = {https://link.aps.org/doi/10.1103/PhysRevLett.100.166401}
}

@article{Nowadnick2017doping,
  title = {Doping dependence of ordered phases and emergent quasiparticles in the doped {H}ubbard-{H}olstein model},
  author = {Mendl, C. B. and Nowadnick, E. A. and Huang, E. W. and Johnston, S. and Moritz, B. and Devereaux, T. P.},
  journal = {Phys. Rev. B},
  volume = {96},
  issue = {20},
  pages = {205141},
  numpages = {7},
  year = {2017},
  month = {Nov},
  publisher = {American Physical Society},
  doi = {10.1103/PhysRevB.96.205141},
  url = {https://link.aps.org/doi/10.1103/PhysRevB.96.205141}
}

@article{Wang2020zerotemperature,
  title = {Zero-temperature phases of the two-dimensional {H}ubbard-{H}olstein model: A non-{G}aussian exact diagonalization study},
  author = {Wang, Yao and Esterlis, Ilya and Shi, Tao and Cirac, J. Ignacio and Demler, Eugene},
  journal = {Phys. Rev. Res.},
  volume = {2},
  issue = {4},
  pages = {043258},
  numpages = {19},
  year = {2020},
  month = {Nov},
  publisher = {American Physical Society},
  doi = {10.1103/PhysRevResearch.2.043258},
  url = {https://link.aps.org/doi/10.1103/PhysRevResearch.2.043258}
}

@article{Frohlich1954electrons, 
    author = {Fr{\"o}hlich, H.},
    year = {1954}, 
    title= {Electrons in Lattice Fields}, 
    journal = {Advances in Physics}, 
    volume={3}, 
    pages={325-361}, 
    url={https://doi.org/10.1080/00018735400101213}
}

@article{MalkarugeCosta2023comparative,
  title = {Comparative determinant quantum {M}onte {C}arlo study of the acoustic and optical variants of the {S}u-{S}chrieffer-{H}eeger model},
  author = {Malkaruge Costa, Sohan and Cohen-Stead, Benjamin and Ly, Andy Tanjaroon and Neuhaus, James and Johnston, Steven},
  journal = {Phys. Rev. B},
  volume = {108},
  issue = {16},
  pages = {165138},
  numpages = {13},
  year = {2023},
  month = {Oct},
  publisher = {American Physical Society},
  doi = {10.1103/PhysRevB.108.165138},
  url = {https://link.aps.org/doi/10.1103/PhysRevB.108.165138}
}

@article{Adolphs2013going,
	author = {Adolphs, C. P. J. and Berciu, M.},
	doi = {10.1209/0295-5075/102/47003},
	journal = {Europhysics Letters},
	month = {jun},
	number = {4},
	pages = {47003},
	publisher = {EDP Sciences, IOP Publishing and Societ{\`a} Italiana di Fisica},
	title = {Going beyond the linear approximation in describing electron-phonon coupling: Relevance for the {H}olstein model},
	url = {https://doi.org/10.1209/0295-5075/102/47003},
	volume = {102},
	year = {2013}}

@article{Ciuchi1997dynamical,
  title = {Dynamical mean-field theory of the small polaron},
  author = {Ciuchi, S. and de Pasquale, F. and Fratini, S. and Feinberg, D.},
  journal = {Phys. Rev. B},
  volume = {56},
  issue = {8},
  pages = {4494--4512},
  numpages = {0},
  year = {1997},
  month = {Aug},
  publisher = {American Physical Society},
  doi = {10.1103/PhysRevB.56.4494},
  url = {https://link.aps.org/doi/10.1103/PhysRevB.56.4494}
}

@article{Paleari2021quantum,
  title = {Quantum {M}onte {C}arlo study of an anharmonic {H}olstein model},
  author = {Paleari, G. and H\'ebert, F. and Cohen-Stead, B. and Barros, K. and Scalettar, RT. and Batrouni, G. G.},
  journal = {Phys. Rev. B},
  volume = {103},
  issue = {19},
  pages = {195117},
  numpages = {11},
  year = {2021},
  month = {May},
  publisher = {American Physical Society},
  doi = {10.1103/PhysRevB.103.195117},
  url = {https://link.aps.org/doi/10.1103/PhysRevB.103.195117}
}

@article{Alexandrov1992polaronic,
  title = {Polaronic effects in the photoemission spectra of strongly coupled electron-phonon systems},
  author = {Alexandrov, A. S. and Ranninger, J.},
  journal = {Phys. Rev. B},
  volume = {45},
  issue = {22},
  pages = {13109(R)--13112(R)},
  numpages = {0},
  year = {1992},
  month = {Jun},
  publisher = {American Physical Society},
  doi = {10.1103/PhysRevB.45.13109},
  url = {https://link.aps.org/doi/10.1103/PhysRevB.45.13109}
}

@article{Ranninger1993spectral,
  title = {Spectral properties of small-polaron systems},
  author = {Ranninger, Julius},
  journal = {Phys. Rev. B},
  volume = {48},
  issue = {17},
  pages = {13166(R)--13169(R)},
  numpages = {0},
  year = {1993},
  month = {Nov},
  publisher = {American Physical Society},
  doi = {10.1103/PhysRevB.48.13166},
  url = {https://link.aps.org/doi/10.1103/PhysRevB.48.13166}
}

@article{Su1979solitons,
  title = {Solitons in Polyacetylene},
  author = {Su, W. P. and Schrieffer, J. R. and Heeger, A. J.},
  journal = {Phys. Rev. Lett.},
  volume = {42},
  issue = {25},
  pages = {1698--1701},
  numpages = {0},
  year = {1979},
  month = {Jun},
  publisher = {American Physical Society},
  doi = {10.1103/PhysRevLett.42.1698},
  url = {https://link.aps.org/doi/10.1103/PhysRevLett.42.1698}
}

@article{Barisic1970tightbinding,
  title = {Tight Binding and Transition-Metal Superconductivity},
  author = {Bari\ifmmode \check{s}\else \v{s}\fi{}i\ifmmode \acute{c}\else \'{c}\fi{}, S. and Labb\'e, J. and Friedel, J.},
  journal = {Phys. Rev. Lett.},
  volume = {25},
  issue = {14},
  pages = {919--922},
  numpages = {0},
  year = {1970},
  month = {Oct},
  publisher = {American Physical Society},
  doi = {10.1103/PhysRevLett.25.919},
  url = {https://link.aps.org/doi/10.1103/PhysRevLett.25.919}
}

@article{Calvani1996polaronic,
  title = {Polaronic optical absorption in electron-doped and hole-doped cuprates},
  author = {Calvani, P. and Capizzi, M. and Lupi, S. and Maselli, P. and Paolone, A. and Roy, P.},
  journal = {Phys. Rev. B},
  volume = {53},
  issue = {5},
  pages = {2756--2766},
  numpages = {0},
  year = {1996},
  month = {Feb},
  publisher = {American Physical Society},
  doi = {10.1103/PhysRevB.53.2756},
  url = {https://link.aps.org/doi/10.1103/PhysRevB.53.2756}
}

@article{Goodvin2011optical,
  title = {Optical Conductivity of the {H}olstein Polaron},
  author = {Goodvin, Glen L. and Mishchenko, Andrey S. and Berciu, Mona},
  journal = {Phys. Rev. Lett.},
  volume = {107},
  issue = {7},
  pages = {076403},
  numpages = {4},
  year = {2011},
  month = {Aug},
  publisher = {American Physical Society},
  doi = {10.1103/PhysRevLett.107.076403},
  url = {https://link.aps.org/doi/10.1103/PhysRevLett.107.076403}
}

@article{Li1999selftrapped,
	author = {Li, Shunran and Luo, Jiajun and Liu, Jing and Tang, Jiang},
	date = {2019/04/18},
	doi = {10.1021/acs.jpclett.8b03604},
	journal = {The Journal of Physical Chemistry Letters},
	journal1 = {The Journal of Physical Chemistry Letters},
	journal2 = {J. Phys. Chem. Lett.},
	month = {04},
	number = {8},
	pages = {1999--2007},
	publisher = {American Chemical Society},
	title = {Self-Trapped Excitons in All-Inorganic Halide Perovskites: Fundamentals, Status, and Potential Applications},
	type = {doi: 10.1021/acs.jpclett.8b03604},
	url = {https://doi.org/10.1021/acs.jpclett.8b03604},
	volume = {10},
	year = {2019},
	year1 = {2019}}

@article{Kandada2020exciton,
	author = {Srimath Kandada, Ajay Ram and Silva, Carlos},
	date = {2020/05/07},
	doi = {10.1021/acs.jpclett.9b02342},
	journal = {The Journal of Physical Chemistry Letters},
	journal1 = {The Journal of Physical Chemistry Letters},
	journal2 = {J. Phys. Chem. Lett.},
	month = {05},
	number = {9},
	pages = {3173--3184},
	publisher = {American Chemical Society},
	title = {Exciton Polarons in Two-Dimensional Hybrid Metal-Halide Perovskites},
	type = {doi: 10.1021/acs.jpclett.9b02342},
	url = {https://doi.org/10.1021/acs.jpclett.9b02342},
	volume = {11},
	year = {2020},
	year1 = {2020}}

@article{Kovac2025Signature,
  title = {Signature of Preformed Pairs in Angle-Resolved Photoemission Spectroscopy},
  author = {Kova\ifmmode \check{c}\else \v{c}\fi{}, Klemen and Nocera, Alberto and Damascelli, Andrea and Bon\ifmmode \check{c}\else \v{c}\fi{}a, Janez and Berciu, Mona},
  journal = {Phys. Rev. Lett.},
  volume = {134},
  issue = {9},
  pages = {096502},
  numpages = {8},
  year = {2025},
  month = {Mar},
  publisher = {American Physical Society},
  doi = {10.1103/PhysRevLett.134.096502},
  url = {https://link.aps.org/doi/10.1103/PhysRevLett.134.096502}
}

@article{Ahn1999Spectral,
  title = {Spectral Evolution in {(Ca,Sr)RuO}$_{3}$ near the {M}ott-{H}ubbard Transition},
  author = {Ahn, J. S. and Bak, J. and Choi, H. S. and Noh, T. W. and Han, J. E. and Bang, Yunkyu and Cho, J. H. and Jia, Q. X.},
  journal = {Phys. Rev. Lett.},
  volume = {82},
  issue = {26},
  pages = {5321--5324},
  numpages = {0},
  year = {1999},
  month = {6},
  publisher = {American Physical Society},
  doi = {10.1103/PhysRevLett.82.5321},
  url = {https://link.aps.org/doi/10.1103/PhysRevLett.82.5321}
}

@article{Esterlis2019pseudogap,
  title = {Pseudogap crossover in the electron-phonon system},
  author = {Esterlis, I. and Kivelson, S. A. and Scalapino, D. J.},
  journal = {Phys. Rev. B},
  volume = {99},
  issue = {17},
  pages = {174516},
  numpages = {5},
  year = {2019},
  month = {May},
  publisher = {American Physical Society},
  doi = {10.1103/PhysRevB.99.174516},
  url = {https://link.aps.org/doi/10.1103/PhysRevB.99.174516}
}

@article{TanjaroonLy2023comparative,
  title = {Comparative study of the superconductivity in the {H}olstein and optical {S}u-{S}chrieffer-{H}eeger models},
  author = {Tanjaroon Ly, Andy and Cohen-Stead, Benjamin and Malkaruge Costa, Sohan and Johnston, Steven},
  journal = {Phys. Rev. B},
  volume = {108},
  issue = {18},
  pages = {184501},
  numpages = {9},
  year = {2023},
  month = {Nov},
  publisher = {American Physical Society},
  doi = {10.1103/PhysRevB.108.184501},
  url = {https://link.aps.org/doi/10.1103/PhysRevB.108.184501}
}

@article{Giustino2017electronphonon,
  title = {Electron-phonon interactions from first principles},
  author = {Giustino, Feliciano},
  journal = {Rev. Mod. Phys.},
  volume = {89},
  issue = {1},
  pages = {015003},
  numpages = {63},
  year = {2017},
  month = {Feb},
  publisher = {American Physical Society},
  doi = {10.1103/RevModPhys.89.015003},
  url = {https://link.aps.org/doi/10.1103/RevModPhys.89.015003}
}

@article{dai2025polarons,
      title={Polarons from first principles}, 
      author={Zhenbang Dai and Jon Lafuente-Bartolome and Feliciano Giustino},
      year={2025},
      journal={arXiv:2512.06176},
      url={https://arxiv.org/abs/2512.06176}, 
}

@article{Salje1994Polarons,
  title={Polarons and bipolarons in tungsten oxide, {WO}$_{3-x}$ },
  author={Ekhard K. H. Salje},
  journal={European Journal of Solid State and Inorganic Chemistry},
  year={1994},
  volume={31},
  pages={805-821},
  url={https://api.semanticscholar.org/CorpusID:101615513}
}

@article{Noack1991CDW,
  title = {Charge-density-wave and pairing susceptibilities in a two-dimensional electron-phonon model},
  author = {Noack, R. M. and Scalapino, D. J. and Scalettar, R. T.},
  journal = {Phys. Rev. Lett.},
  volume = {66},
  issue = {6},
  pages = {778--781},
  numpages = {0},
  year = {1991},
  month = {Feb},
  publisher = {American Physical Society},
  doi = {10.1103/PhysRevLett.66.778},
  url = {https://link.aps.org/doi/10.1103/PhysRevLett.66.778}
}

@article{Niyaz1993CDW,
  title = {Charge-density-wave-gap formation in the two-dimensional {H}olstein model at half-filling},
  author = {Niyaz, Parhat and Gubernatis, J. E. and Scalettar, R. T. and Fong, C. Y.},
  journal = {Phys. Rev. B},
  volume = {48},
  issue = {21},
  pages = {16011--16022},
  numpages = {0},
  year = {1993},
  month = {Dec},
  publisher = {American Physical Society},
  doi = {10.1103/PhysRevB.48.16011},
  url = {https://link.aps.org/doi/10.1103/PhysRevB.48.16011}
}

@article{Alder1997Variational,
  title = {Variational {M}onte {C}arlo Study of an Interacting Electron-Phonon Model},
  author = {Alder, B. J. and Runge, K. J. and Scalettar, R. T.},
  journal = {Phys. Rev. Lett.},
  volume = {79},
  issue = {16},
  pages = {3022--3025},
  numpages = {0},
  year = {1997},
  month = {Oct},
  publisher = {American Physical Society},
  doi = {10.1103/PhysRevLett.79.3022},
  url = {https://link.aps.org/doi/10.1103/PhysRevLett.79.3022}
}

@article{Sykora2009Coexistence,
doi = {10.1209/0295-5075/85/57003},
url = {https://dx.doi.org/10.1209/0295-5075/85/57003},
year = {2009},
month = {mar},
publisher = {},
volume = {85},
number = {5},
pages = {57003},
author = {Sykora, S. and Hübsch, A. and Becker, K. W.},
title = {Coexistence of superconductivity and charge-density waves in a two-dimensional {H}olstein model at half-filling},
journal = {Europhysics Letters}
}

@article{Weber2018TwoD,
  title = {Two-dimensional {H}olstein-{H}ubbard model: Critical temperature, {I}sing universality, and bipolaron liquid},
  author = {Weber, Manuel and Hohenadler, Martin},
  journal = {Phys. Rev. B},
  volume = {98},
  issue = {8},
  pages = {085405},
  numpages = {8},
  year = {2018},
  month = {Aug},
  publisher = {American Physical Society},
  doi = {10.1103/PhysRevB.98.085405},
  url = {https://link.aps.org/doi/10.1103/PhysRevB.98.085405}
}

@article{Araujo2022TwoD,
  title = {Two-dimensional $t\ensuremath{-}{t}^\prime$ {H}olstein model},
  author = {Ara\'ujo, Maykon V. and de Lima, Jos\'e P. and Sorella, Sandro and Costa, Natanael C.},
  journal = {Phys. Rev. B},
  volume = {105},
  issue = {16},
  pages = {165103},
  numpages = {8},
  year = {2022},
  month = {Apr},
  publisher = {American Physical Society},
  doi = {10.1103/PhysRevB.105.165103},
  url = {https://link.aps.org/doi/10.1103/PhysRevB.105.165103}
}

@article{Ostmeyer2025Minimal,
title = {Minimal autocorrelation in hybrid {M}onte {C}arlo simulations using exact {F}ourier acceleration},
journal = {Computer Physics Communications},
volume = {313},
pages = {109624},
year = {2025},
issn = {0010-4655},
doi = {https://doi.org/10.1016/j.cpc.2025.109624},
url = {https://www.sciencedirect.com/science/article/pii/S0010465525001262},
author = {Johann Ostmeyer and Pavel Buividovich}
}

@article{Kaufmann2023anacont,
title = {ana\_cont: {P}ython package for analytic continuation},
journal = {Computer Physics Communications},
volume = {282},
pages = {108519},
year = {2023},
issn = {0010-4655},
doi = {https://doi.org/10.1016/j.cpc.2022.108519},
url = {https://www.sciencedirect.com/science/article/pii/S0010465522002387},
author = {Josef Kaufmann and Karsten Held},
}

@article{Bradley2021Superconductivity,
  title = {Superconductivity and charge density wave order in the two-dimensional {H}olstein model},
  author = {Bradley, Owen and Batrouni, George G. and Scalettar, Richard T.},
  journal = {Phys. Rev. B},
  volume = {103},
  issue = {23},
  pages = {235104},
  numpages = {9},
  year = {2021},
  month = {Jun},
  publisher = {American Physical Society},
  doi = {10.1103/PhysRevB.103.235104},
  url = {https://link.aps.org/doi/10.1103/PhysRevB.103.235104}
}

@software{PaperRepo1,
  author       = {Neuhaus, James},
  title        = {Scripts and Plots for {H}olstein {S}pectral and {C}onductive {P}roperties paper},
  month        = jul,
  year         = 2026,
  publisher    = {Zenodo},
  version      = {v0.1},
  doi          = {10.5281/zenodo.21221268},
  url          = {https://doi.org/10.5281/zenodo.21221268},
  howpublished = {https://doi.org/10.5281/zenodo.21221268},
}

@misc{PaperRepo2,
  author       = {Neuhaus, James},
  title        = {Full dataset and visualizers for {H}olstein {S}pectral and {C}onductive {P}roperties paper},
  month        = jul,
  year         = 2026,
  publisher    = {Zenodo},
  doi          = {10.5281/zenodo.21220924},
  url          = {https://doi.org/10.5281/zenodo.21220924},
  howpublished = {https://doi.org/10.5281/zenodo.21220924},
}

@article{Moser2013Tunable,
  title = {Tunable Polaronic Conduction in Anatase {$\mathrm{TiO}_{2}$}},
  author = {Moser, S. and Moreschini, L. and Ja\ifmmode \acute{c}\else \'{c}\fi{}imovi\ifmmode \acute{c}\else \'{c}\fi{}, J. and Bari\ifmmode \check{s}\else \v{s}\fi{}i\ifmmode \acute{c}\else \'{c}\fi{}, O. S. and Berger, H. and Magrez, A. and Chang, Y. J. and Kim, K. S. and Bostwick, A. and Rotenberg, E. and Forr\'o, L. and Grioni, M.},
  journal = {Phys. Rev. Lett.},
  volume = {110},
  issue = {19},
  pages = {196403},
  numpages = {5},
  year = {2013},
  month = {May},
  publisher = {American Physical Society},
  doi = {10.1103/PhysRevLett.110.196403},
  url = {https://link.aps.org/doi/10.1103/PhysRevLett.110.196403}
}

\end{document}